\documentclass[journal]{IEEEtran}  %

\usepackage[table]{xcolor} %
\usepackage{graphicx}

\usepackage{multirow}      %
\usepackage{booktabs}      %
\usepackage{amsmath,amssymb,amsfonts}
\usepackage{algorithm}
\usepackage{algorithmic}
\usepackage{subfigure}
\usepackage{cite}

\usepackage{mdframed}
\usepackage{enumitem}

\usepackage{fontawesome5}
\newcommand{\revised}[1]{{\color{black}#1}}

\begin{document}

\bstctlcite{IEEEexample:BSTcontrol}
\title{PENDA: An Efficient Processing Element via Norm-of-Difference for Deep Learning Accelerators}

\author{\IEEEauthorblockN{Kai-Chieh Hsu, and Tian-Sheuan Chang, \textit{Senior Member, IEEE}}
\thanks{This work was supported by the National Science and Technology Council, Taiwan, under Grant 111-2622-8-A49-018-SB, 110-2221-E-A49-148-MY3, 113-2221-E-A49-078-MY3, and 113-2640-E-A49-005. The authors are affiliated with the Institute of Electronics, National Yang Ming Chiao Tung University, Taiwan. (e-mail: hsukaij.ee11@nycu.edu.tw, tschang@nycu.edu.tw) }%
\thanks{Manuscript received XXXX XX, 2025; revised XXXX XX, XXXX.}
}
\maketitle

\begin{abstract}%
Inner product computation dominates the computational cost of deep learning models; thus, accelerating this primitive is key to improving hardware efficiency. However, most existing techniques rely on approximations, which can degrade model accuracy. To preserve exactness while optimizing hardware, this paper presents PENDA (processing element via norm-of-difference architecture), which leverages the law of cosines to recast multiplications as squared-difference operations. Replacing multiply–accumulate units with the proposed norm-of-difference units yields 11–36\%, 5–48\%, and 11–19\% reductions in area, energy, and clock period, respectively, for the PE array of a deep learning accelerator.

\end{abstract}

\section{Introduction}
\label{chapter:introduction}

The rapid advancement of deep learning has produced increasingly complex models, escalating the demand for specialized hardware accelerators such as ASICs and FPGAs. For the core of these models, general matrix multiplication (GEMM) is the dominant computational workload, serving as a fundamental building block in both Transformer architectures~\cite{NIPS2017_3f5ee243} and convolutional neural networks (CNNs)~\cite{726791}. Conventional deep learning accelerators typically employ processing elements (PEs) based on multiply–accumulate (MAC) units to execute these operations. 
\revised{However, the multiplier within a MAC unit remains an important contributor to silicon area, energy consumption, and critical-path delay. Therefore, improving the efficiency of PE-level inner-product computation is critical for deep learning accelerators.}

\revised{

Table~\ref{table: comparison with baseline and the previous works} summarizes the benefits and limitations of representative prior techniques for improving PE-array efficiency. Sparsity-based accelerators~\cite{qin2020sigma, hsu2025low} improve performance and energy efficiency by skipping zero or pruned operands. However, their benefits depend on the availability of sufficient sparsity, and aggressive weight pruning generally requires fine-tuning to recover accuracy. Approximate-computing approaches~\cite{schonleber2025stella, liu2024encodingnet} can provide substantial efficiency gains by replacing exact multiplications with lower-cost approximate operations. Nevertheless, these methods trade arithmetic exactness for efficiency, have mainly been evaluated on CNN inference workloads, and often require additional fine-tuning, thereby limiting their generality across model families and usage scenarios. To avoid accuracy loss, exact algebraic methods such as Winograd’s Fast Inner Product (FIP) and Free-Pipeline Fast Inner Product (FFIP)~\cite{1687427,pogue2023fast} preserve numerical correctness while reducing the number of multiplications. However, these approaches still rely on multiplier-based datapaths and are tightly coupled to specific PE-array organizations, thereby limiting their flexibility across diverse accelerator dataflows. These limitations motivate a more general PE-level solution that can improve hardware efficiency, preserve exact computation, and remain compatible with diverse accelerator architectures.

}

\begin{table}[]
\caption{\revised{comparison with the previous works} }
\centering

\newcommand{\gsmile}{\raisebox{-0.3ex}{\textcolor{green}{\faIcon[regular]{smile}}}}
\newcommand{\rfrown}{\raisebox{-0.3ex}{\textcolor{red}{\faIcon[regular]{frown}}}}

\begin{tabular}{cccccc} 
\toprule
                                                                &   MAC  & \begin{tabular}[c]{@{}c@{}} Sparse \\ ~\cite{qin2020sigma, hsu2025low}\end{tabular} & \begin{tabular}[c]{@{}c@{}} Approx. \\ ~\cite{schonleber2025stella, liu2024encodingnet} \end{tabular} & \begin{tabular}[c]{@{}c@{}}  Winograd \\ FIP \\ ~\cite{1687427,pogue2023fast} \end{tabular} & Ours \\ 
\midrule
\rule[-2ex]{0pt}{4.2ex} Clock Rate                                         & \rfrown & \rfrown & \gsmile & \rfrown & \gsmile \\  \hline
\begin{tabular}[c]{@{}c@{}}Area\\ Efficiency\end{tabular}       & \rfrown & \gsmile & \gsmile & \gsmile & \gsmile \\  \hline
\begin{tabular}[c]{@{}c@{}}Energy per\\ Operation\end{tabular}  & \rfrown & \gsmile & \gsmile & \gsmile & \gsmile \\  \hline
\rule[-2ex]{0pt}{5ex} Exactness                                          & \gsmile & \rfrown & \rfrown & \gsmile & \gsmile \\  \hline
\begin{tabular}[c]{@{}c@{}}Model\\ Generality\end{tabular}      & \gsmile & \rfrown & \rfrown & \gsmile & \gsmile \\  \hline
\begin{tabular}[c]{@{}c@{}}Dataflow \\ Flexibility\end{tabular} & \gsmile & \rfrown & \rfrown & \rfrown & \gsmile \\  \hline
\begin{tabular}[c]{@{}c@{}}Fine-tuning \\ Overhead\end{tabular} & \gsmile & \rfrown & \rfrown & \gsmile & \gsmile \\  \hline
\bottomrule
\end{tabular}
\label{table: comparison with baseline and the previous works}
\end{table}

\begin{figure}[tbp]
\centering
\includegraphics[height=!,width=1.0\linewidth,keepaspectratio=true]{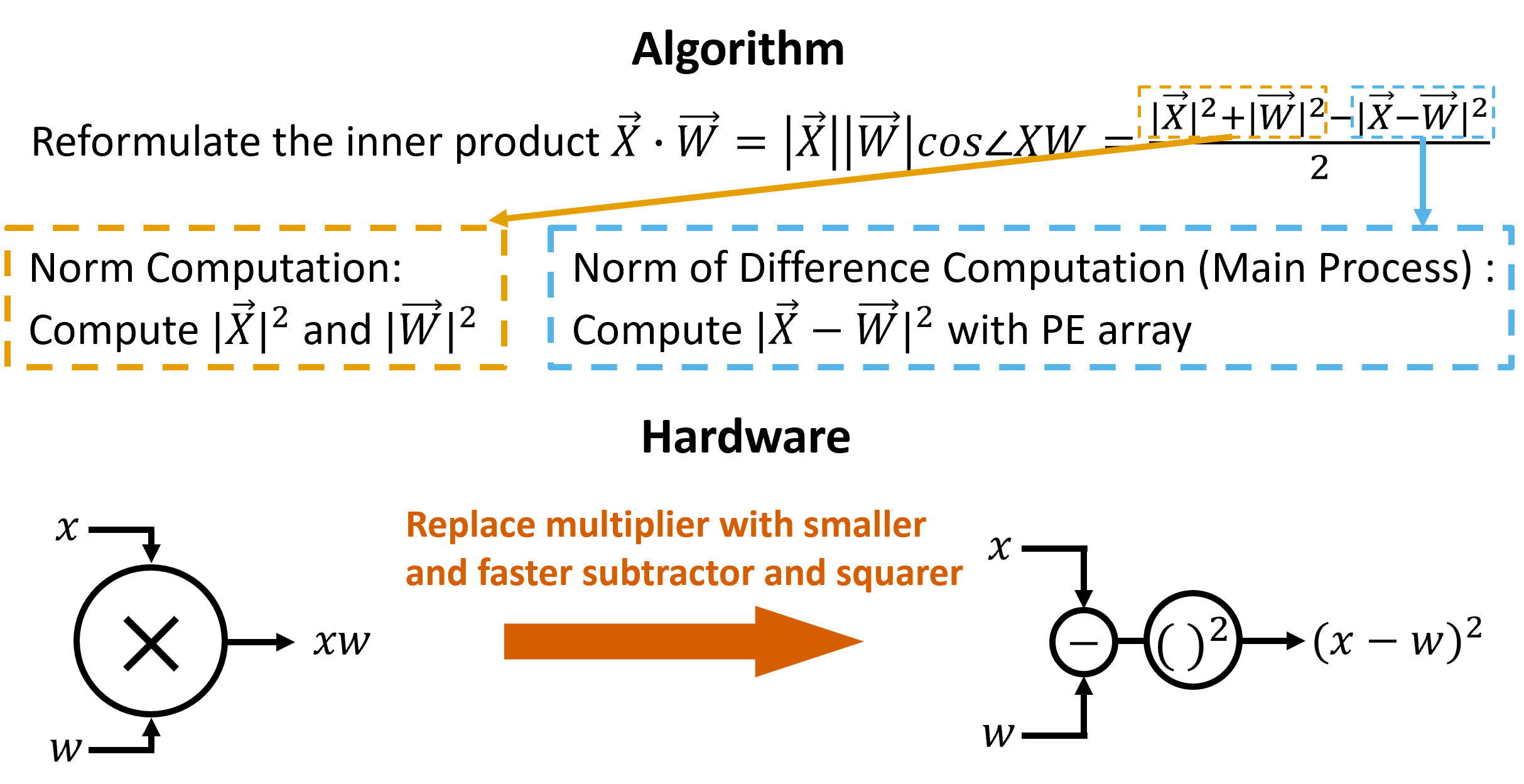}
\caption{Overview of proposed method}
\label{fig overview}
\end{figure}

In this paper, we propose PENDA (processing element via norm-of-difference architecture), as illustrated in Fig.~\ref{fig overview}. This architecture leverages the law of cosines to algebraically transform the conventional MAC operation into a hardware-efficient norm-of-difference (NoD) operation, thereby replacing the standard MAC with the proposed NoD unit in deep learning accelerators. By eliminating the traditional MAC and adopting NoD, PENDA reduces PE area and energy consumption with negligible overhead and no loss of computational accuracy. Notably, PENDA performs operations using efficient squarers instead of conventional multipliers, enabling a higher clock frequency due to the reduced arithmetic complexity. Moreover, implementing PENDA requires modifications almost exclusively within the PE, allowing integration with a wide variety of existing PE array architectures.


The remainder of the paper is organized as follows. Section~II reviews the related work on inner product acceleration. Section~III details the proposed PENDA methodology. Section~IV describes the hardware implementation. Experimental results are provided in Section~V, followed by concluding remarks in Section~VI.

\section{Related Work}
\label{chapter:Related Work}

In the domain of deep learning hardware acceleration, the efficiency of inner product computations is a critical design consideration, as these operations form the computational foundation of all major model layers. Traditional accelerators rely on MAC, which is often among the major contributors to chip area and power consumption. To mitigate this bottleneck, researchers have explored algorithmically efficient alternatives.

\subsection{Sparsity-Based, Approximate, and Quantized Methods}
A variety of approaches have been explored to reduce the computational burden of PEs. First, pruning and sparsity-aware computation, exemplified by SIGMA~\cite{qin2020sigma}, SparTen~\cite{sparten} and our prior work~\cite{hsu2025low}, reduce energy by exploiting data sparsity. Second, hash-based product quantization, such as Maddness~\cite{blalock2021multiplying} and Stella Nera~\cite{schonleber2025stella}, replaces multiplications with simple table lookups and additions, thereby avoiding direct MAC operations. Furthermore, coding-based MAC designs~\cite{liu2024encodingnet} simplify multipliers through wide-bit projections and bit-weighted accumulations. While these strategies improve area and energy efficiency, the approximate methods trade arithmetic exactness for efficiency, and the sparsity-based methods depend on pruning that generally requires fine-tuning to recover accuracy. Moreover, their applicability across diverse deep learning models is often limited.

\subsection{Winograd-Based Exact Methods}
To preserve exact arithmetic, another influential line of research centers on algebraic re-formulation. A foundational approach here is the FIP algorithm~\cite{1687427}, introduced by S. Winograd in 1968. The central principle of FIP is an algebraic transformation, as shown in \eqref{eq:FIP}, that allows matrix multiplication to be performed with approximately half the number of MAC operations. This significant reduction is achieved by replacing computationally expensive multiplications with a greater quantity of lower-cost, low-bitwidth additions.

\begin{equation}
\label{eq:FIP}
c_{i,j} = \sum_{k=1}^{K/2} \bigl(a_{i,2k-1}+b_{2k,j}\bigr)
          \bigl(a_{i,2k}+b_{2k-1,j}\bigr) - \alpha_i - \beta_j ,
\end{equation}
where
\begin{equation*}
\alpha_i = \sum_{j=1}^{K/2} a_{i,2j-1}\,a_{i,2j},
\qquad
\beta_j  = \sum_{i=1}^{K/2} b_{2i-1,j}\,b_{2i,j}.
\end{equation*}

An important property of this formulation is that the input pre-transformation term, \(\alpha\) depends only on the input matrix A and is completely independent of B, while the kernel pre-transformation term, \(\beta\), depends only on B. As a result, \(\alpha\) and \(\beta\) can be computed independently and their values can be extensively reused across multiple output computations, making the initial overhead for calculating \(\alpha\) and \(\beta\) negligible. Consequently, the dominant computation time is spent in evaluating the leading summation term of the equation. From a theoretical standpoint, this trade-off presents a compelling path to improving the compute efficiency and performance per unit area of a hardware accelerator. However, the FIP algorithm has remained largely under-explored for decades due to a significant practical drawback: its hardware implementation results in a notable reduction in clock frequency, which consequently decreases overall throughput. This limitation has historically undermined the algorithm's theoretical benefits. Pogue and Nicolici~\cite{pogue2023fast} address this critical weakness with their FFIP algorithm. Their innovation lies in a new hardware architecture that enables the FFIP to achieve the same nearly 2$\times$ reduction in MAC units while inherently improving clock frequency and throughput, making the algorithm viable for modern, high-performance deep learning accelerators. Nevertheless, FFIP still faces two fundamental constraints: it is tightly coupled to a specific systolic-array architecture, and it continues to depend on MAC-based computation, leaving its maximum clock frequency bounded by multiplier latency. These limitations underscore the need for a more general solution that eliminates general-purpose multipliers, which motivates the work presented in this paper.
\section{Proposed Method}

\subsection{Overview of PENDA}
\label{Overview of DISCOS}

The foundational principle of PENDA is an algebraic reformulation of the inner product computation. Instead of conventional MAC, PENDA leverages the law of cosines to transform the operation. In a standard matrix multiplication \(\mathbf{Y} = \mathbf{X}\mathbf{W}\), each element \(y_{ij}\) represents the inner product of a row vector \(\vec{X_i}\) from \(\mathbf{X}\) and a column vector \(\vec{W_j}\) from \(\mathbf{W}\). By applying the law of cosines, this inner product \(\vec{X_i} \cdot \vec{W_j}\) can be re-expressed using squared Euclidean norms, as derived in \eqref{cos law}.

\begin{align}
\label{cos law}
&y_{ij} = \vec{X_i} \cdot \vec{W_j} \nonumber\\
&= |\vec{X_i}||\vec{W_j}| \cos\angle \vec{X_i} \vec{W_j} \nonumber\\
&= |\vec{X_i}||\vec{W_j}| \times \frac{|\vec{X_i}|^2 + |\vec{W_j}|^2 - |\vec{X_i} - \vec{W_j}|^2}{2|\vec{X_i}||\vec{W_j}|} \nonumber\\
&= \frac{|\vec{X_i}|^2 + |\vec{W_j}|^2 - |\vec{X_i} - \vec{W_j}|^2}{2},
\end{align}

where
    
\begin{equation*}
|\vec{X_i}|^2  = \sum_{k=1}^{K} x_{ik}^2,
\qquad
|\vec{W_j}|^2  = \sum_{k=1}^{K} w_{kj}^2,
\end{equation*}

\begin{equation*}
|\vec{Xi}-\vec{Wj}|^2  = \sum_{k=1}^{K} (x_{ik}-w_{kj})^2
\end{equation*}

This transformation effectively converts the entire matrix multiplication from a MAC-based problem into one solvable by norm and NoD operations.

A critical property of \eqref{cos law} is the high reusability of the norm terms, \(|\vec{X_i}|^2\) and \(|\vec{W_j}|^2\). The term \(|\vec{X_i}|^2\) is computed once per row of \(\mathbf{X}\) and reused for all $N$ columns, while $|\vec{W_j}|^2$ is computed once per column of $\mathbf{W}$ and reused for all $M$ rows. Consequently, the dominant computational burden of the operation shifts from the original MAC operations to the evaluation of the NoD term, $|\vec{Xi}-\vec{Wj}|^2$. This computational shift from MAC to NoD computation provides significant hardware advantages. 

\begin{figure}[bp]
\subfigure[Partial products to be reduced in a multiplier.]{
    \includegraphics[width=0.4\linewidth]{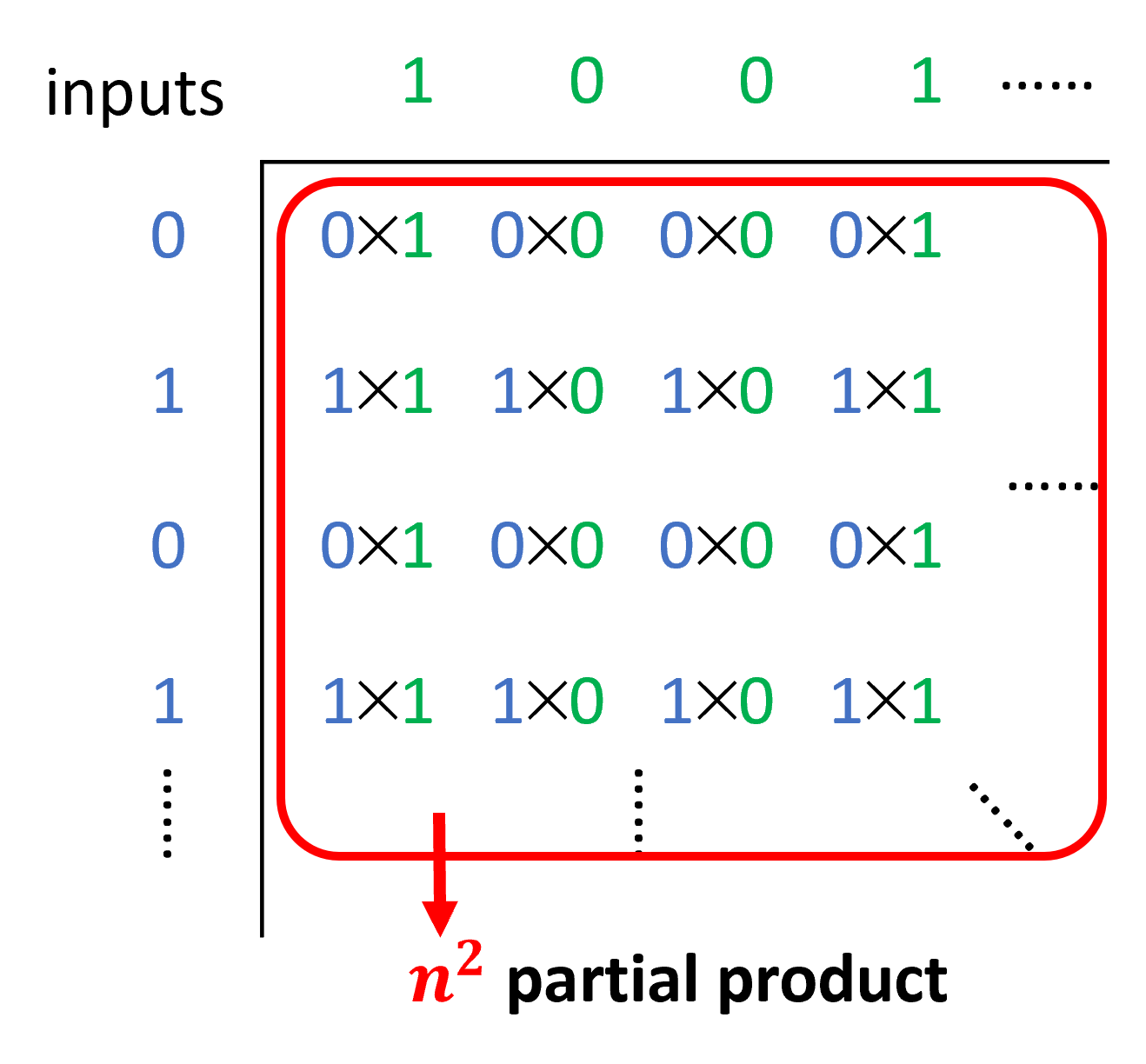}
    \label{fig:1a}
}
\subfigure[Partial products to be reduced in a squarer.]{
    \includegraphics[width=0.4\linewidth]{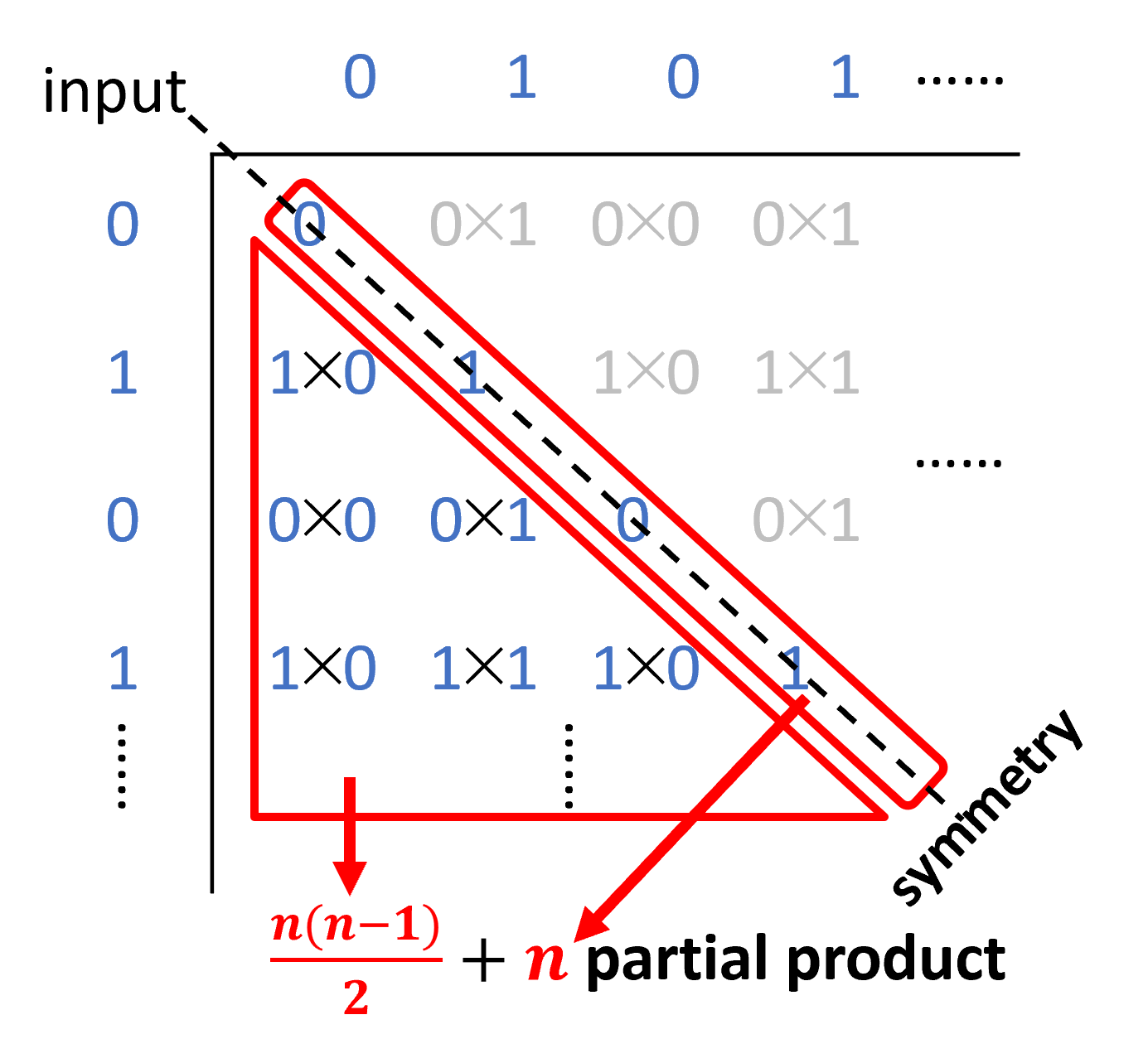}
    \label{fig:1c}
}
\caption{The internal architecture of the multiplier and squarer.}
\label{mult vs square}
\end{figure}

The proposed PENDA replaces the complex multiplier with a simpler subtractor and a squarer. A conventional n-bit multiplier's structure involves generating $n^2$ partial products, which are then combined, as illustrated in Fig.~\ref{mult vs square}. In contrast, a squarer exhibits symmetric partial products, requiring only about half as many to be generated and combined. This simpler reduction tree also leads to a shorter critical path, enabling a higher clock frequency (i.e., a shorter clock period). Consequently, the area and power consumption of a squarer are approximately half those of a multiplier. Leveraging this property, PENDA achieves substantial reductions in both area and power consumption while preserving full computational accuracy.

\begin{algorithm}[!t]
\caption{Computation of $Y$ from $X$ and $W$}
\label{alg:DISCOS}
\begin{algorithmic}
\REQUIRE $X \in \mathbb{R}^{M\times K}$, $W \in \mathbb{R}^{K\times N}$

\STATE \textbf{\textit{Offline Compute:}}
\FOR{$j = 1$ to $N$}
    \STATE $Wnorm_j \gets \displaystyle \sum_{k=1}^{K} w_{k j}^2$
\ENDFOR

\STATE \textbf{\textit{Stage 1 - NoD Computation:}}
\FOR{$i = 1$ to $M$}
    \FOR{$j = 1$ to $N$}
        \STATE $D_{i j} \gets \displaystyle \sum_{k=1}^{K} \bigl(x_{i k} - w_{k j}\bigr)^2$
    \ENDFOR
\ENDFOR

\STATE \textbf{\textit{Stage 2 - Norm Computation:}}
\FOR{$i = 1$ to $M$}
    \STATE $Xnorm_i \gets \displaystyle \sum_{k=1}^{K} x_{i k}^2$
\ENDFOR

\STATE \textbf{\textit{Stage 3 - Final Combination:}}
\FOR{$i = 1$ to $M$}
    \FOR{$j = 1$ to $N$}
        \STATE $y_{i j} \gets \dfrac{Xnorm_i + Wnorm_j - D_{i j}}{2}$
    \ENDFOR
\ENDFOR

\end{algorithmic}
\end{algorithm}

Algorithm~\ref{alg:DISCOS} details the procedure for performing matrix multiplication using PENDA, illustrated with the pseudo-code example of computing $\mathbf{Y} = \mathbf{X}\mathbf{W}$. The procedure consists of three stages.

The primary computational step is the NoD computation. According to \eqref{cos law}, this stage computes $|\vec{Xi}-\vec{Wj}|^2$, that is, $\sum_{k=1}^{K} (x_{ik}-w_{kj})^2$, which consists of first computing the difference, then squaring, and finally accumulating the results. The NoD computation is executed by the modified NoD-based PE array, which will be described in detail in Section~\ref{Proposed DSAC}. This stage's dataflow, input pattern, and iteration order are identical to those of conventional MAC-based matrix multiplication.


The second stage is the norm computation, which computes the norms $|\vec{X_i}|^2$ and $|\vec{W_j}|^2$. The details are described in Section~\ref{Norm-Square Computation}. This process is handled differently for weights and activations. The norm of the weights, $|\vec{W_j}|^2$, can be pre-computed offline, incurring no runtime cost. The norm of the activations, $|\vec{X_i}|^2$, must be computed online. This online computation can be executed concurrently with the NoD computation by utilizing a portion of the PE array resources. This parallel execution, combined with high data reusability, ensures the overhead of this online stage accounts for a negligible fraction of the total computational cost.

The last stage is the final combination. In this step, the results from the NoD computation and the norm computation (i.e., $|\vec{X_i}|^2$, $|\vec{W_j}|^2$, and $|\vec{Xi}-\vec{Wj}|^2$) are combined through simple addition and subtraction as specified by \eqref{cos law} to reconstruct the final result.


\subsection{NoD Computation}
\label{Proposed DSAC}
\begin{figure}[tbp]
\centering
\includegraphics[height=!,width=0.6\linewidth,keepaspectratio=true]{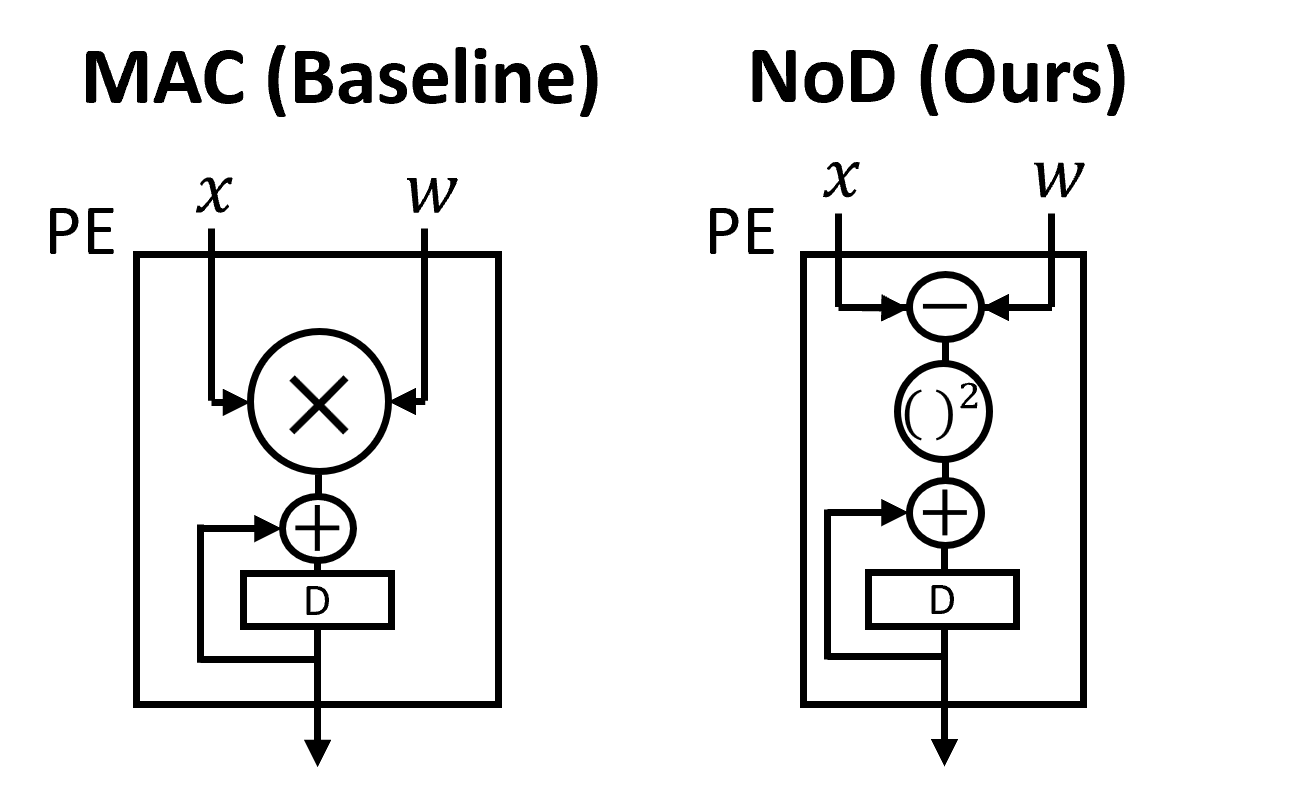}
\caption{Comparison between conventional MAC-based PE and NoD-based PE.}
\label{MAC_vs_DSAC}
\end{figure}
The NoD-based PE is the core component that executes the NoD computation. Each PE integrates a subtractor, a squarer, and an accumulator, as illustrated in Fig.~\ref{MAC_vs_DSAC}, a design that occupies less on-chip area and consumes less energy than a conventional MAC while achieving a higher maximum clock frequency. An array of these PEs forms the computational core, where each PE executes the subtract–square–accumulate sequence on its operands to compute the primary difference term, $|\vec{Xi}-\vec{Wj}|^2$.

\subsection{Norm Computation}
\label{Norm-Square Computation}


The norm computation stage calculates $|\vec{X_i}|^2$ and $|\vec{W_j}|^2$. Since the NoD computation serves as the direct architectural counterpart to the baseline MAC workload (substituting it with identical complexity), these auxiliary norm calculations constitute the method's overhead. Weight norms $|\vec{W_j}|^2$ are precomputed offline and fused with biases, with no runtime overhead. However, activation norms $|\vec{X_i}|^2$ are computed online and buffered. To minimize the overhead of this online computation, we propose three implementation strategies whose execution schedules are illustrated in Fig.~\ref{fig:schedule}:

\begin{figure}[t]
\centering
\includegraphics[height=!,width=1.0\linewidth,keepaspectratio=true]{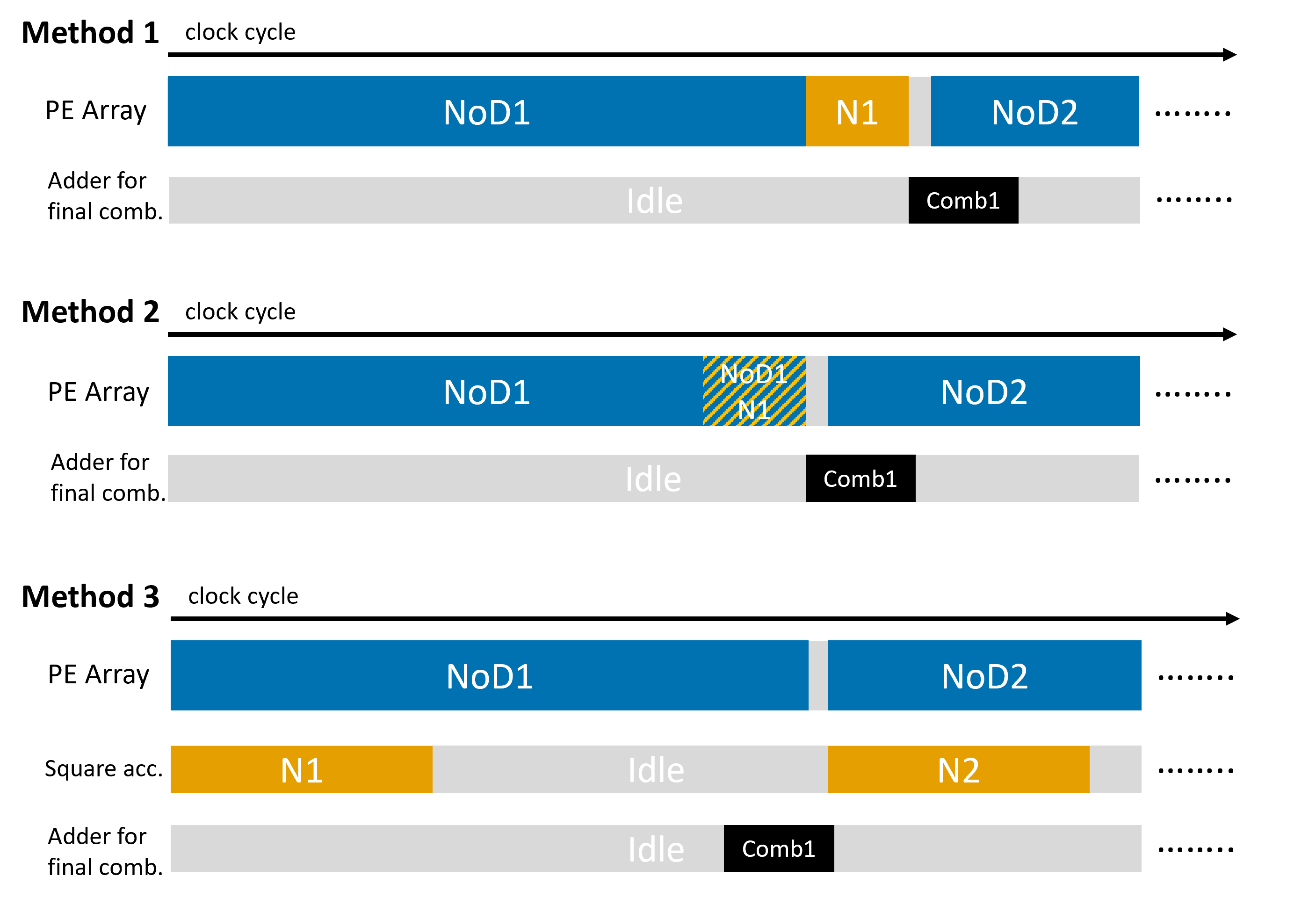}
\caption{Schematic diagram of the system execution schedules for different implementation methods. Method 1: Norm computation is performed sequentially, incurring approx. 1.6\% execution-time overhead, while PEs are underutilized during the NoD stage. Method 2: The PE array concurrently performs NoD and norm computations by utilizing naturally idle PEs. Method 3: Dedicated square accumulators compute norms concurrently while the main PE array performs NoD computations. 'N' denotes norm computation, 'Comb' denotes final combination, and the suffix number indicates the matrix operation instance.}
\label{fig:schedule}
\end{figure}

\begin{figure}[t]
\centering
\includegraphics[height=!,width=0.98\linewidth,keepaspectratio=true]{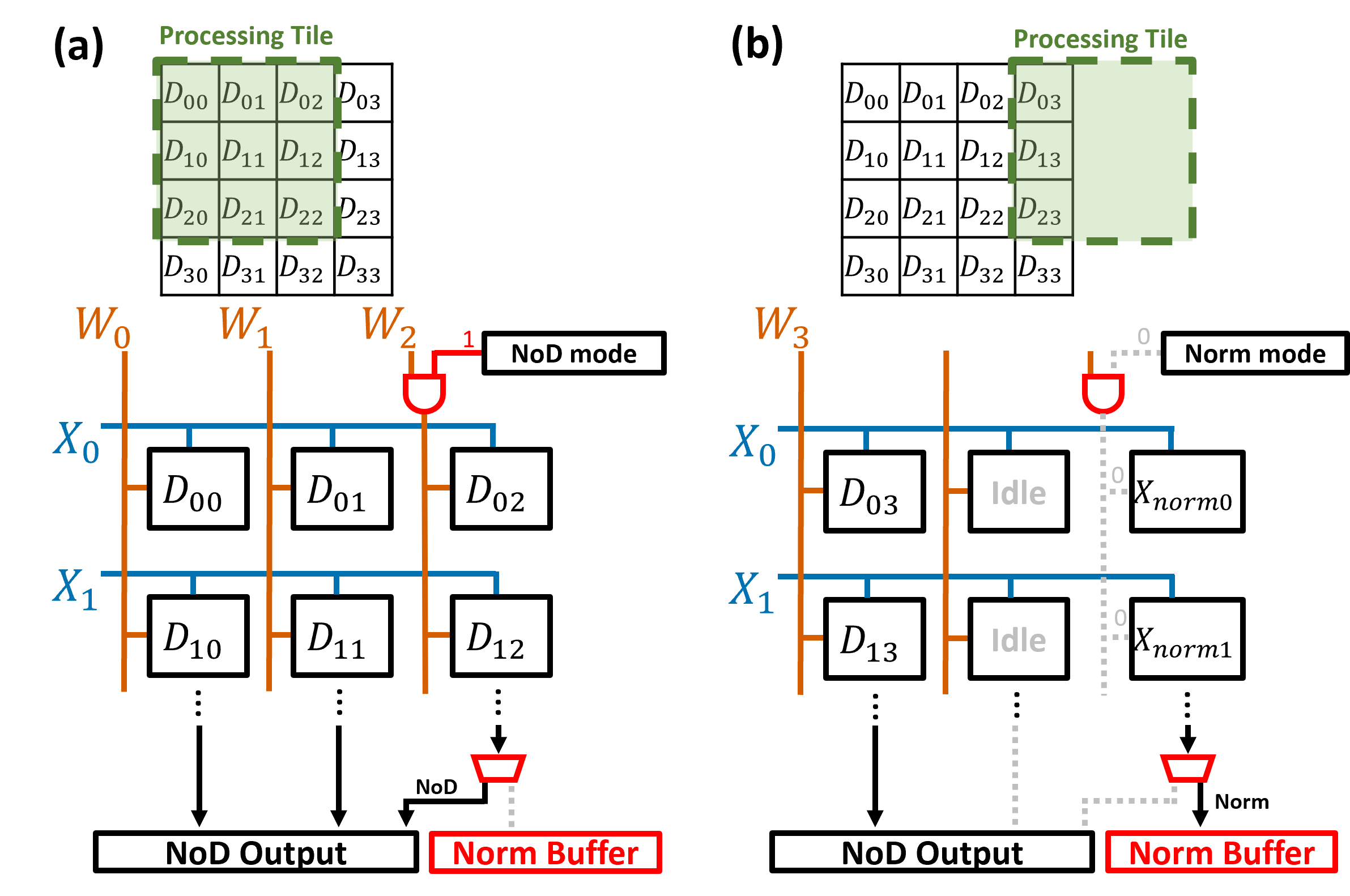}
\caption{\revised{Norm computation using the idle last PE lane. (a) A non-boundary tile fully occupies the PE array, and the last PE lane remains in normal NoD mode. (b) A boundary tile leaves the last PE lane idle, allowing it to compute activation norms. Red blocks denote the additional hardware for Method 2.}}
\label{norm_method2}
\end{figure}

\begin{figure}[htbp]
\centering
\includegraphics[height=!,width=0.85\linewidth,keepaspectratio=true]{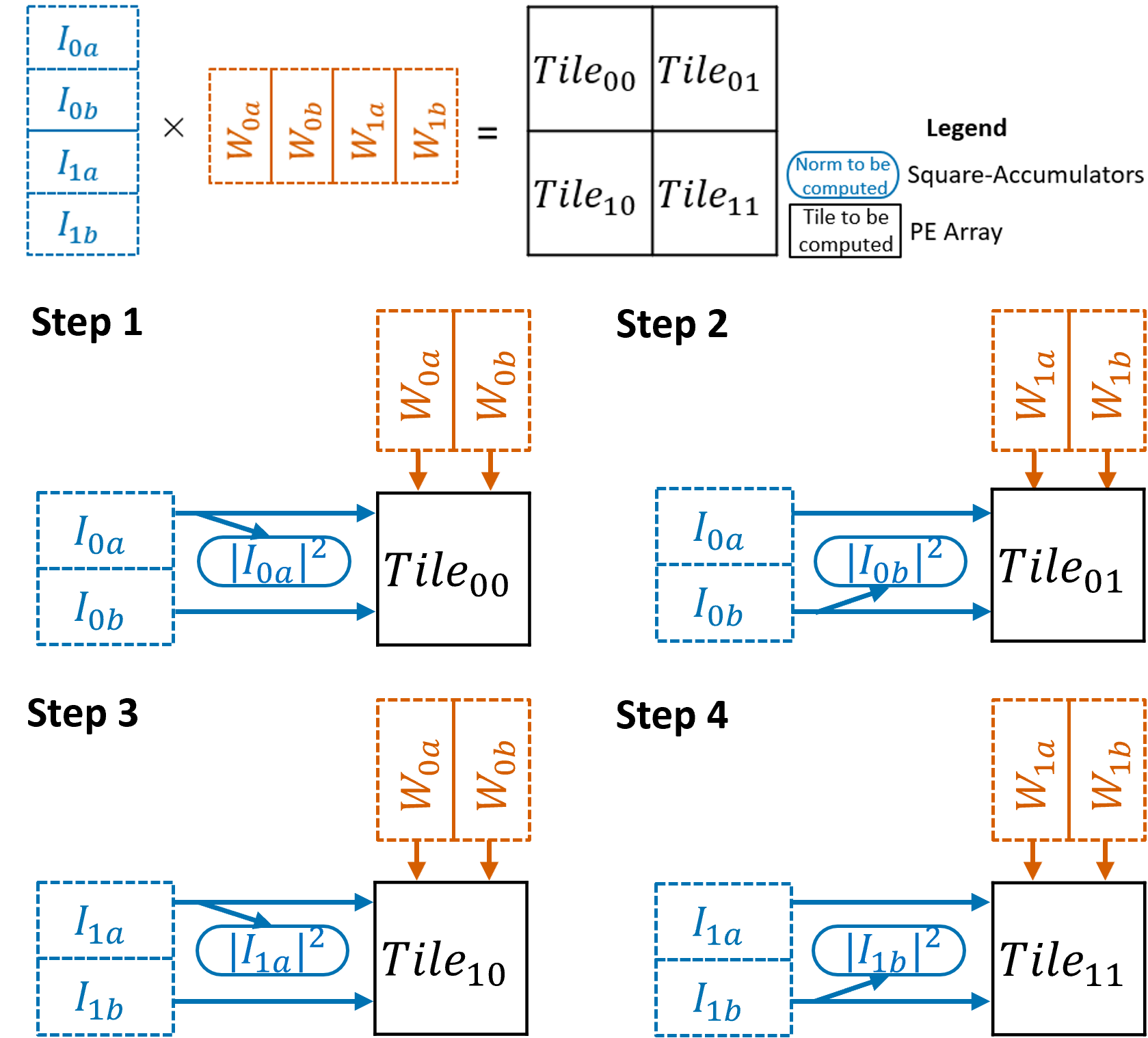}
\caption{Norm computation employs dedicated square accumulators.}
\label{Norm_Square_SAC}
\end{figure}

\begin{itemize}
\item Method 1 (Sequential): The PE array calculates norms sequentially after the NoD term. This requires no hardware modification but incurs execution time overhead.
\item Method 2 (Concurrent Execution via Idle PEs): To eliminate time overhead, we exploit the partial utilization of PE inherent in tiling-based scheduling. 
\revised{
Since matrix dimensions are often not exact multiples of the PE-array size, some PEs naturally remain
idle when processing boundary tiles. As illustrated in Fig.~\ref{norm_method2}, the last PE lane is reused to compute activation norms concurrently. In norm-computation mode, a small AND gate forces the weight input of this lane to zero. A demultiplexer at the output of the last PE lane then redirects the result to the norm buffer instead of the normal NoD output path.}
\item Method 3 (Dedicated Hardware): \revised{As illustrated in Fig.~\ref{Norm_Square_SAC},} an alternative approach employs dedicated square accumulators that operate concurrently with the main PE array. These accumulators tap into the NoD data stream during data fetching, incurring no additional buffer read bandwidth. By leveraging the repetitive data fetches inherent in tiled matrix multiplication, they process different portions of the input data across passes until all norms $|\vec{X_i}|^2$ are computed. This approach eliminates latency overhead at the cost of a small area increase.
\end{itemize}

\subsection{Overhead Analysis}
\label{sec:Overhead_Analysis}

\begin{table}[htbp]

\caption{\revised{Computational Overhead for a ViT-B Encoder Layer}}

\centering
\begin{tabular}{lccc} 
    \toprule 
    Linear Operation & \begin{tabular}[c]{@{}c@{}}Baseline Ops \\ ($A$)\end{tabular} & \begin{tabular}[c]{@{}c@{}} Norm Ops \\ ($B$)\end{tabular} & \begin{tabular}[c]{@{}c@{}} Overhead \\  ($B/A$)\end{tabular} \\
    \midrule 
    
    $XW_{Q}/XW_{K}/XW_{V}$ & 349M & 0.15M & 0.04\% \\
    QK                     & 30M  & 0.3M  & 1.02\% \\
    AV                     & 30M  & 0.62M & 2.07\% \\
    $ZW_{O}$               & 116M & 0.15M & 0.13\% \\
    $\mathrm{FFN_{1}}$              & 465M & 0.15M & 0.03\% \\
    $\mathrm{FFN_{2}}$              & 465M & 0.61M & 0.13\% \\
    
    \midrule 
    Total                  & 1454M & 1.98M & 0.14\% \\
    \bottomrule 
\end{tabular}
\\[1.0ex]
\footnotesize
\parbox{\linewidth}{
\textit{Note:} Workload is reported in PE Operations to preclude ambiguity (2 ops for MAC vs. 3 ops for NoD). One PE Operation is defined as the atomic processing of a single input pair.  \revised{Offline precomputed weight norms are excluded.}
}

\label{table:MACs_overhead}
\end{table}

The norm computation introduces potential sources of overhead in computation, storage, and system control. Here, we evaluate these costs using a ViT-B~\cite{dosovitskiy2020vit} encoder layer as a benchmark.

\subsubsection{Computational Overhead} \revised{Norm calculation requires $M \times K$ operations, which is asymptotically negligible compared to the $M \times N \times K$ core workload of the NoD computation and its baseline MAC counterpart.} Quantitatively, Table~\ref{table:MACs_overhead} confirms that norm operations account for only 0.14\% of the total baseline operations.

\subsubsection{Execution Time Overhead} In our benchmark, Method 1 incurs a 1.6\% cycle overhead due to sequential execution. In contrast, Method 2 eliminates this overhead by optimizing the dataflow to utilize idle PEs, effectively masking the latency. Alternatively, Method 3 guarantees zero time overhead via dedicated hardware at a marginal cost relative to the PE array (2.5\% area, 0.6\% power).

\subsubsection{Storage Overhead} Buffering activation norms requires only $O(M)$ space, a small fraction of the $O(M \times K)$ activation storage. Table~\ref{table:Storage_overhead} confirms this overhead is negligible, peaking at 1.56\% and typically remaining below 1\%.

\subsubsection{Control and Dataflow Overhead} A key advantage of PENDA is that the core NoD computation strictly adheres to the dataflow, input pattern, and iteration order of the conventional MAC baseline. Consequently, the internal control logic and data movement of the baseline array remain unchanged except for the small mode-control and output-routing logic. At the system level, modifications are minimal and confined to peripheral operations: specifically, managing the temporary buffering of activation norms and executing the final combination step before output. Method 1 may also require an additional input read phase. 

Overall, these overheads are small relative to the measured performance gains and do not dominate the evaluated system cost or latency.

\begin{table}[]

\caption{Storage Overhead for a ViT-B Encoder Layer}

\centering
\begin{tabular}{cccc}
    \toprule 
    
    \begin{tabular}[c]{@{}c@{}}Input Act. \end{tabular} & 
    \begin{tabular}[c]{@{}c@{}}Activation Size\\ ($A$) \end{tabular} & 
    \begin{tabular}[c]{@{}c@{}}Norm Size\\ ($B$) \end{tabular} & 
    \begin{tabular}[c]{@{}c@{}}Storage Overhead\\ ($B/A$) \end{tabular} \\
    
    \midrule 
    
    $X_{in}$   & 151K & 197  & 0.13\% \\
    Q/K/V      & 454K & 7092 & 1.56\% \\
    A          & 466K & 2364 & 0.51\% \\
    Z          & 151K & 197  & 0.13\% \\
    $X_{\mathrm{FFN1}}$ & 151K & 197  & 0.13\% \\
    $X_{\mathrm{FFN2}}$ & 605K & 197  & 0.03\% \\
    
    \bottomrule 
\end{tabular}

\label{table:Storage_overhead}
\end{table}

\section{Hardware Design for Deep Learning Accelerators}
\label{chapter:Hardware}

The following presents how to apply the proposed design to deep learning accelerators, including PE designs, a specialized PENDA-CNN design, and a TPU-style accelerator~\cite{jouppi2017datacenter},~\cite{dao2022flashattention}.

\subsection{PE}
\label{sec:PE}
\begin{figure}[bp]
\centering
\subfigure[FP16]{%
    \includegraphics[width=0.8\linewidth]{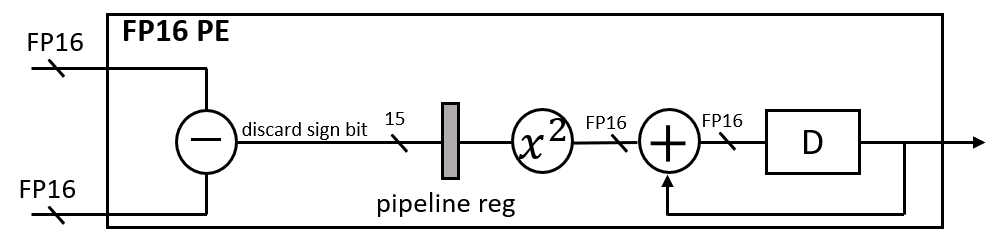}\label{Detailed_PE_FP16}}
\\
\subfigure[INT16]{%
    \includegraphics[width=0.8\linewidth]{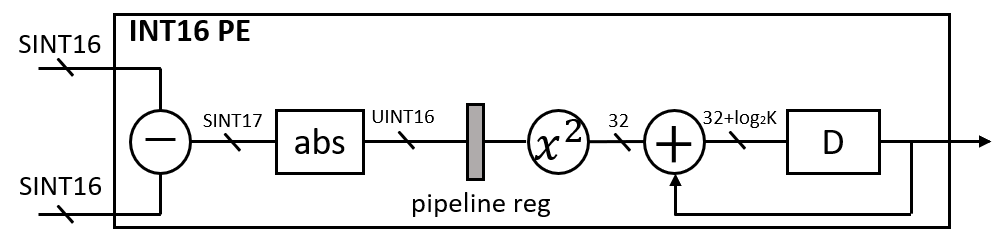}\label{Detailed_PE_INT16}}
\caption{\revised{Detailed diagram of a PE. SINT and UINT denote signed and unsigned integer, respectively.}}
\label{Detailed_PE}
\end{figure}

Fig.~\ref{Detailed_PE} shows the detailed NoD-based PEs for FP16 and INT16, each of which consists of a subtractor, a squarer, an accumulator, and a single pipeline register.
This pipeline stage is introduced to achieve a substantial increase in PE clock frequency at a negligible area cost. For example, under the INT16 configuration, pipelining raises the clock frequency by 30\% while requiring only one additional 16-bit register, resulting in an area increase of only 5\%. Compared with a conventional MAC-based PE, the proposed NoD-based PE achieves a 20\% speedup. In contrast, traditional MAC-based designs are constrained by the multiplier-dominated critical path; further acceleration typically necessitates inserting multiple pipeline stages within the multiplier, which incurs a large number of pipeline registers.

A critical design consideration for integer-based PEs is the accumulator bit-width required to prevent overflow. The accumulation value range of NoD (i.e., $\sum (x_{ik}-w_{kj})^2$) differs from that of a conventional MAC ($\sum x_{ik} \cdot w_{kj}$). We can theoretically analyze this difference by modeling the inputs $A$ and $B$ as independent, identically distributed Gaussian random variables with zero mean ($\mu=0$) and standard deviation $\sigma$. Under these assumptions, the standard deviation of the MAC term is $\sigma_{MAC} = \sigma(A \cdot B)$, while the standard deviation of the NoD term is $\sigma_{NoD} = \sigma((A-B)^2)$. The theoretical ratio of these standard deviations is $\sigma_{NoD} / \sigma_{MAC} = 2\sqrt{2}$. This ratio ($\approx 2.83$) implies that the value range of the NoD accumulation is larger, suggesting a need for approximately 2 additional bits in the accumulator to provide the same headroom against overflow as a MAC unit. 

We validated this estimation empirically using the ViT-B~\cite{dosovitskiy2020vit} model. For INT8 operations, the accumulator of MAC required 24 bits to prevent overflow, whereas the accumulator of NoD required 26 bits. Similarly, for INT16 operations, the  accumulators of MAC and NoD required 40 and 42 bits, respectively. These experimental results, consistently showing a +2 bit requirement for NoD, align with our theoretical analysis.

\revised{We also analyzed the ResNet-18~\cite{he2016deep}. The required accumulator width is 20 bits for the MAC-based PE and 23 bits for the NoD-based PE. This is slightly larger than the Gaussian-based theoretical +2-bit estimate, which may result from the stronger spatial correlation of CNN feature maps and their larger deviation from the independent Gaussian assumption used in the theoretical analysis.

Therefore, in our hardware implementation, the NoD-based PE accumulator is provisioned with approximately two to three additional bits of precision depending on the target workload.
}

\begin{figure}[tbp]
\centering
\includegraphics[height=!,width=1.0\linewidth,keepaspectratio=true]{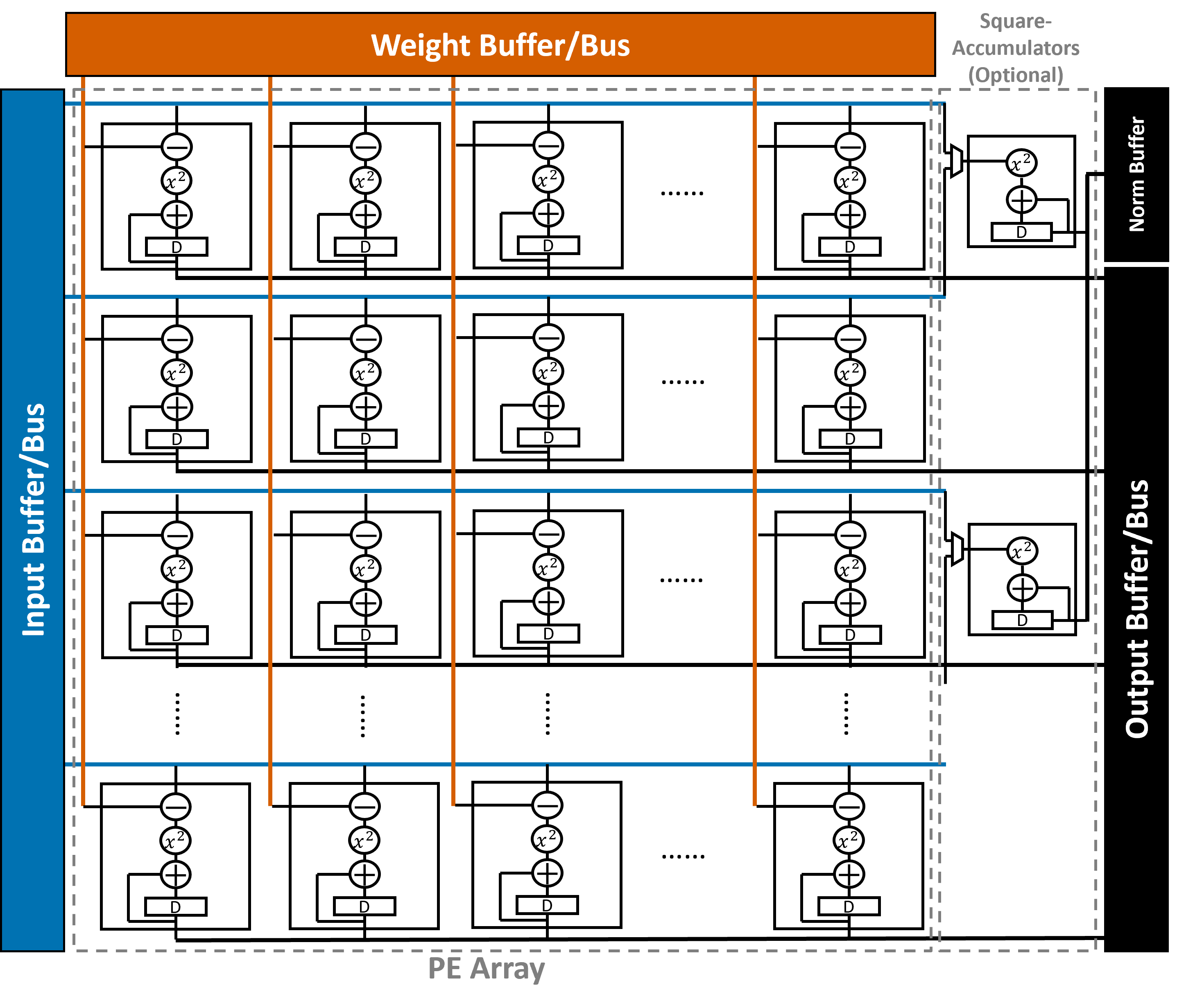}
\caption{Detailed diagram of an output-stationary NoD-based PE array with optional square-accumulators.}
\label{OSSA}
\end{figure}

\subsection{PE Array}
\label{sec:PE Array}

\begin{figure}[t]
\centering
\includegraphics[height=!,width=1.0\linewidth,keepaspectratio=true]{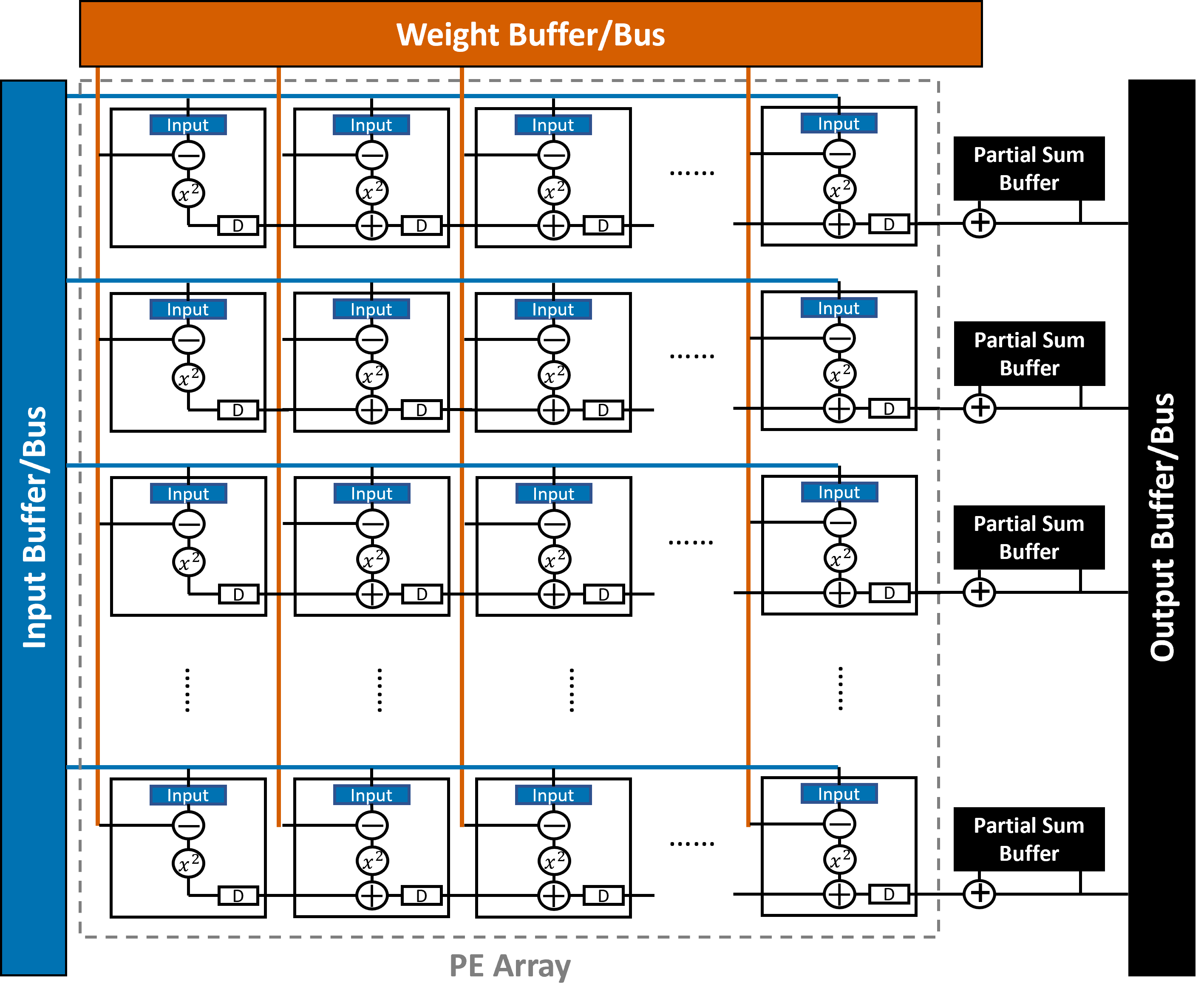}
\caption{Detailed diagram of an input-stationary NoD-based PE array.}
\label{ISSA}
\end{figure}

Based on the proposed NoD-based PE, we realize two PE-array organizations: output-stationary (OS) and input-stationary (IS) (configurable to weight-stationary). Fig.~\ref{OSSA} shows the NoD-based OS array. Fig.~\ref{ISSA} illustrates the NoD-based IS array. These complementary dataflows enable the runtime to select a mapping that maximizes data reuse under given layer shapes and bandwidth constraints.


\begin{figure}[tbp]
\centering
\includegraphics[height=!,width=0.8\linewidth,keepaspectratio=true]{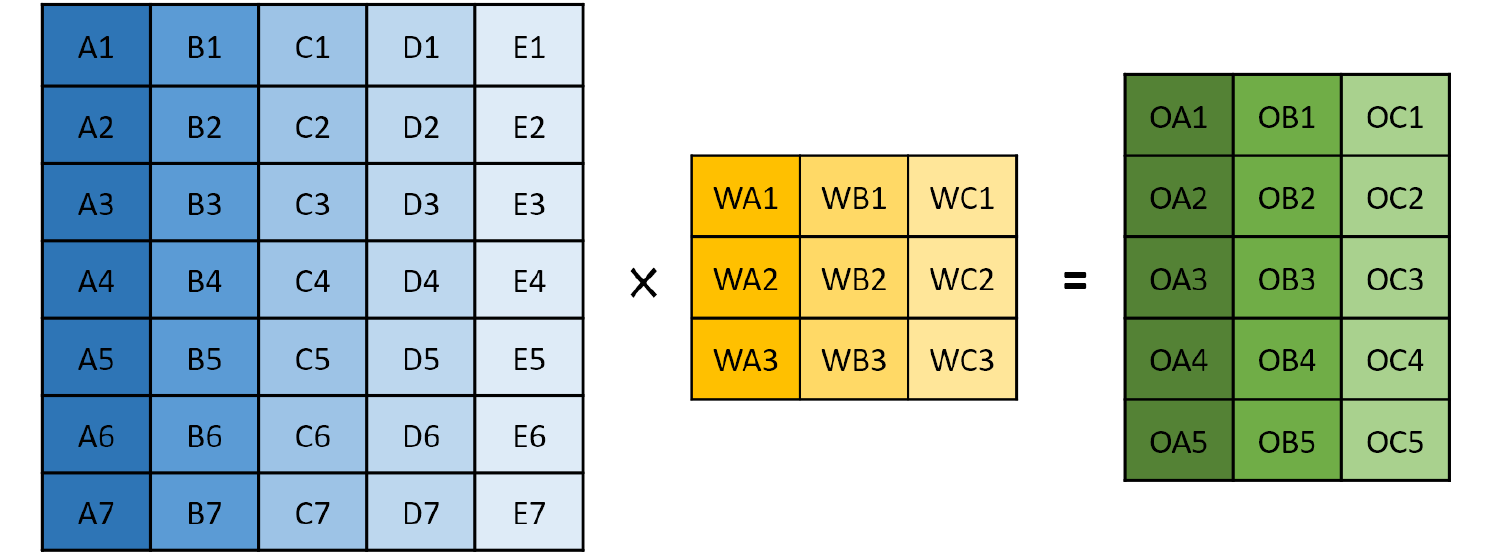}
\caption{An example of CNN operation.}
\label{CNN_example}
\vspace{0.7cm}
\centering
\includegraphics[height=!,width=1.0\linewidth,keepaspectratio=true]{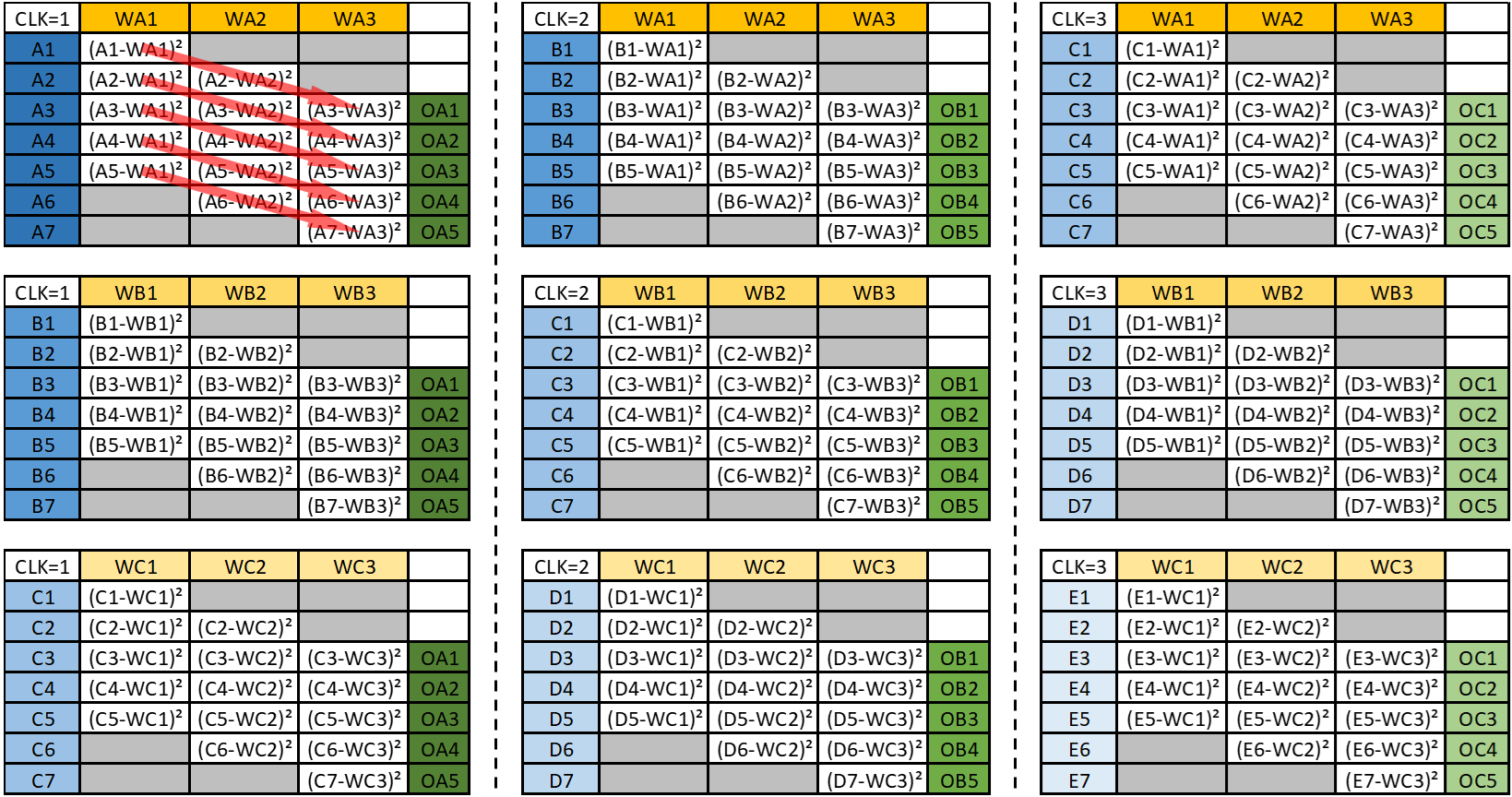}
\caption{Dataflow of PENDA-CNN.}
\label{DISCOS_CNN}
\end{figure}

\subsection{PENDA-CNN}
\label{sec:DISCOS CNN}

Beyond the general-purpose OS and IS array configurations, PENDA is also applicable to CNN acceleration. By integrating PENDA with the CNN-based PE architecture described in~\cite{chang2019vwa},~\cite{huang2022real}, we propose the architecture termed PENDA-CNN. 

Fig.~\ref{CNN_example} presents an example of the convolution operation. Correspondingly, Fig.~\ref{DISCOS_CNN} illustrates the dataflow of PENDA-CNN that executes this operation. The input feature map \(\mathbf{X}\) and the filter \(\mathbf{W}\) are partitioned column-wise. At each step, a column from the input feature map and a row from the filter are selected. These selected data undergo the NoD computation. The outputs are then accumulated diagonally to generate the partial sum of NoD (i.e., \(|\vec{Xi}-\vec{Wj}|^2\) ) corresponding to the selected pair of feature map column and filter row. By aggregating these corresponding partial sums, the complete NoD for the entire convolution operation is obtained.

\begin{figure}[htbp]
\centering
\includegraphics[height=!,width=1.0\linewidth,keepaspectratio=true]{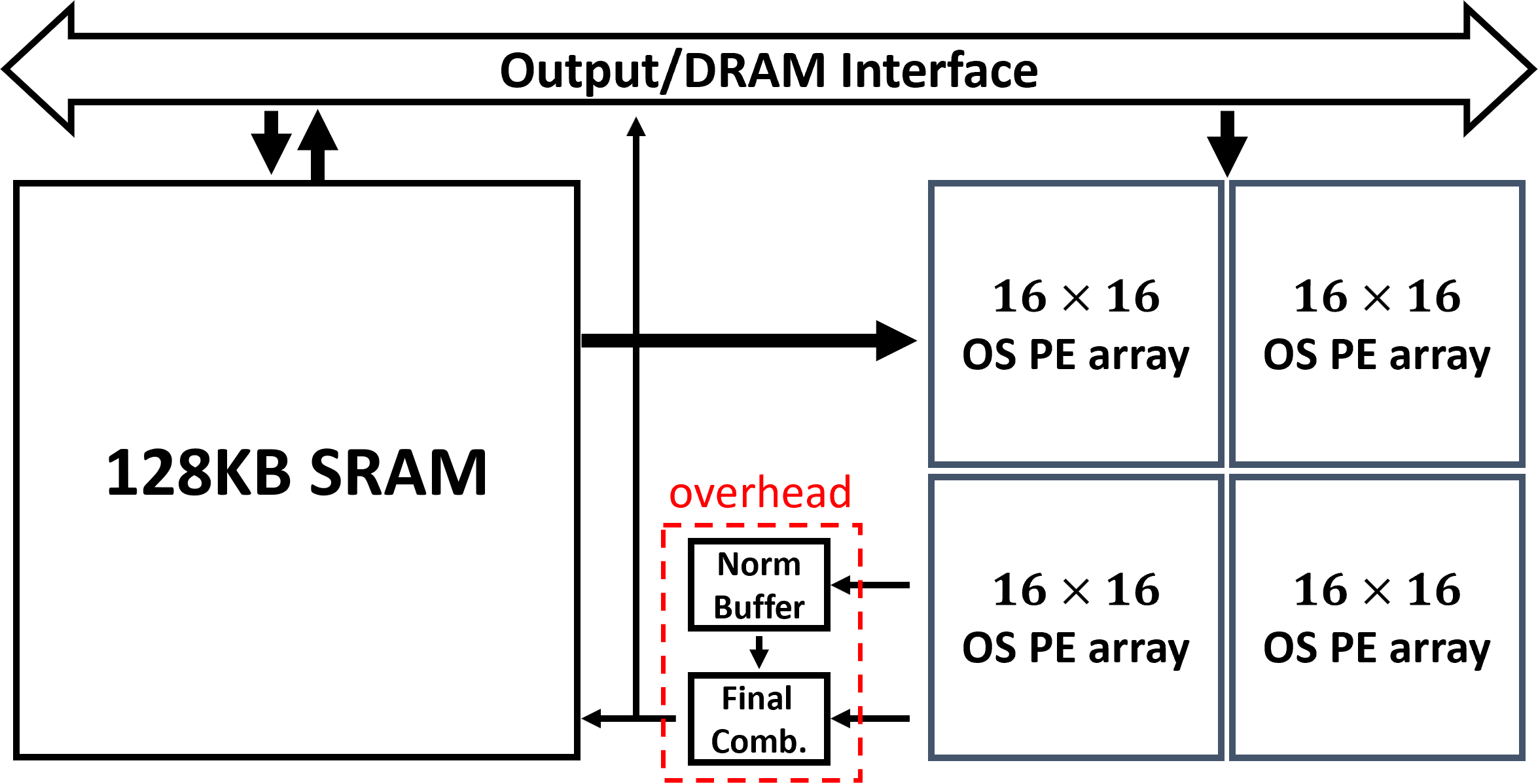}
\caption{System Architecture.}
\label{fig:System}
\end{figure}

\begin{table*}[t]

\caption{Comparison Across Various Precisions on a $16\times16$ OS PE Array}

\centering
\begin{tabular}{cccccccccc}
    \toprule 
    
    Precision & PE type & 
    \begin{tabular}[c]{@{}c@{}}Area \\ ($\mu\mathrm{m}^{2}$)\end{tabular} & 
    \begin{tabular}[c]{@{}c@{}}Area\\ Reduction\\ (vs. MAC)\end{tabular} & 
    \begin{tabular}[c]{@{}c@{}}Power\\ (mW)\end{tabular} & 
    \begin{tabular}[c]{@{}c@{}}Normalized\\ Energy\\ ($\mathrm{mW\cdot ns}$)\end{tabular} & 
    \begin{tabular}[c]{@{}c@{}}Energy\\ Reduction\\ (vs. MAC)\end{tabular} & 
    \begin{tabular}[c]{@{}c@{}}Clock \\ Period \\ (ns)\end{tabular} & 
    \begin{tabular}[c]{@{}c@{}}Speedup\\ (vs. MAC)\end{tabular} & 
    \begin{tabular}[c]{@{}c@{}}AET \\ Reduction \\ (vs. MAC)\end{tabular} \\ 
    
    \midrule 

    \multirow{2}{*}{INT8} & MAC & 44607 & - & 180 & 90 & - & 0.5 & - & - \\
                          & NoD & 38010 & 15\% & 194 & 85 & 5\% & 0.44 & 14\% & 29\% \\ 
    \midrule 

    \multirow{2}{*}{INT16}& MAC & 108639 & - & 376 & 241 & - & 0.64 & - & - \\
                          & NoD & 96385 & 11\% & 402 & 217 & 10\% & 0.54 & 19\% & 33\% \\ 
    \midrule 

    \multirow{2}{*}{FP16} & MAC & 219334 & - & 501 & 801 & - & 1.6 & - & - \\
                          & NoD & 139575 & 36\% & 309 & 417 & 48\% & 1.35 & 19\% & 72\% \\ 
    \midrule 

    \multirow{2}{*}{FP32} & MAC & 494384 & - & 854 & 1708 & - & 2 & - & - \\
                          & NoD & 361685 & 27\% & 673 & 1211 & 29\% & 1.8 & 11\% & 53\% \\ 
    
    \bottomrule 
\end{tabular}
\\[1.0ex]
\footnotesize
\parbox{\textwidth}{
\textit{Note 1:} All integer accumulator bit-widths are provisioned to be overflow-free based on the analysis in Section \ref{sec:PE} (INT8: 24-bit MAC vs. 26-bit NoD; INT16: 40-bit MAC vs. 42-bit NoD).

\textit{Note 2:} Normalized Energy = Power $\times$ Clock Period
}

\label{tab:os_comparison}
\end{table*}

\subsection{System Architecture}
\label{sec:System Architecture}

To evaluate system-level performance, we define a TPU-style accelerator architecture~\cite{jouppi2017datacenter}, illustrated in Fig. \ref{fig:System}. The system is composed of a 128 kB SRAM buffer and a computational core consisting of four 16$\times$16 INT8 OS PE arrays, which operate in parallel. When implementing the baseline MAC-based version, the 128 kB SRAM serves as the buffer. 
\revised{For the proposed PENDA version, an additional 1.6 kB SRAM is required to store the norms of runtime-generated intermediate variables, along with adders for the final combination step. The norm buffer stores one norm value for each tiled input row, with size \(S_{\text{norm}}=N_{\text{op}}\times N_{\text{norm}}\times B_{\text{norm}}\), where \(N_{\text{op}}\) is the number of runtime operands requiring online norms, \(N_{\text{norm}}\) is the number of norm values buffered per operand, and \(B_{\text{norm}}\) is the byte width of each norm value. 
For ViT-B, norms are stored per tiled input row, giving $N_{\text{norm}}=16\lceil197/16\rceil$. With \(N_{\text{op}}=2\) and \(B_{\text{norm}}=4\), the norm buffer requires 1664 bytes \(\approx\) 1.6 kB. 

We also evaluate a ResNet-18 CNN accelerator~\cite{chang2019vwa,huang2022real}. The system instantiates \(64\times3\) groups of \(5\times3\) row-wise convolution PE arrays, corresponding to \(64\times5\times3\times3\) PEs for channel, pixel, and \(3\times3\) kernel parallelism. It uses a 128 kB buffer and applies layer fusion to shallow layers. The norm buffer stores one norm value per feature-map pixel. Excluding the tiled shallow layers, the maximum feature-map size is $28\times28$; with each unquantized 23-bit norm stored in 3 bytes, the required norm buffer is $1\times(28\times28)\times3$ bytes, approximately 2.3 kB.
}

\section{Experimental Results}
\label{chapter:Experimental Result}

In this section, we evaluate the hardware performance of the proposed PENDA. We first conduct a direct component-level comparison between our NoD-based PE and the conventional MAC-based PE within identical OS and IS array configurations. We then benchmark our PENDA against a state-of-the-art FFIP~\cite{pogue2023fast} architecture, highlighting a critical implementation bottleneck in prior work. Finally, we present a system-level comparison based on a TPU-style accelerator to demonstrate the end-to-end performance gains.

To obtain concrete hardware metrics for our component-level and state-of-the-art comparisons, all PE array designs (MAC, NoD, and FFIP) were synthesized using the TSMC 16-nm process. We used Synopsys Design Compiler to derive area metrics and Synopsys PrimeTime for power analysis. All designs were synthesized under identical conditions to ensure a fair comparison.

\subsection{Component-Level Comparison (NoD vs. MAC)}
\label{Quantization}

The objective of this first comparison is to demonstrate the fundamental hardware benefits of replacing MAC units with NoD units. We synthesized 16x16 PE arrays for both OS and IS dataflows, ensuring all parameters were identical, with the only variable being the PE core (MAC vs. NoD).

Table \ref{tab:os_comparison} presents the results for the OS array. As indicated by the data, the NoD-based PE achieves consistent reductions in area, energy, and clock period across all tested precisions (INT8, INT16, FP16, and FP32). Across the tested precisions, the NoD-based PE results in 11--36\% reduction in area, \mbox{5--48\%} in energy, and 11--19\% in clock period, culminating in a significant 29--72\% reduction in the composite Area-Energy-Time (AET) metric. Regarding power consumption, a direct comparison is inequitable due to the differing clock periods; the NoD-based PE supports a higher operating frequency, which naturally elevates dynamic power. Therefore, Normalized Energy (Power $\times$ Clock Period) serves as the fair metric, representing the energy consumption per clock cycle. This normalization is strictly valid because the dataflows of PENDA and the conventional MAC-based accelerators are identical, and the execution cycle overhead is negligible. Under this equitable metric, the NoD-based PE demonstrates superior energy efficiency compared to the MAC baseline.

\revised{
At low precision, the benefit of PENDA becomes less pronounced because the baseline multiplier is already small and the additional subtractor overhead becomes more visible. For INT4, PENDA still reduces area by 13\%, but its energy increases by 13\%, leading to a smaller AET reduction of 9\%. For FP8 and lower FP formats, the cost of floating-point subtraction can offset the savings from replacing multiplication with squaring. Therefore, PENDA is most beneficial for INT8, FP16, and higher-precision configurations.
}

\begin{figure}[tbp]
\centering
\includegraphics[width=1.0\linewidth, keepaspectratio=true]{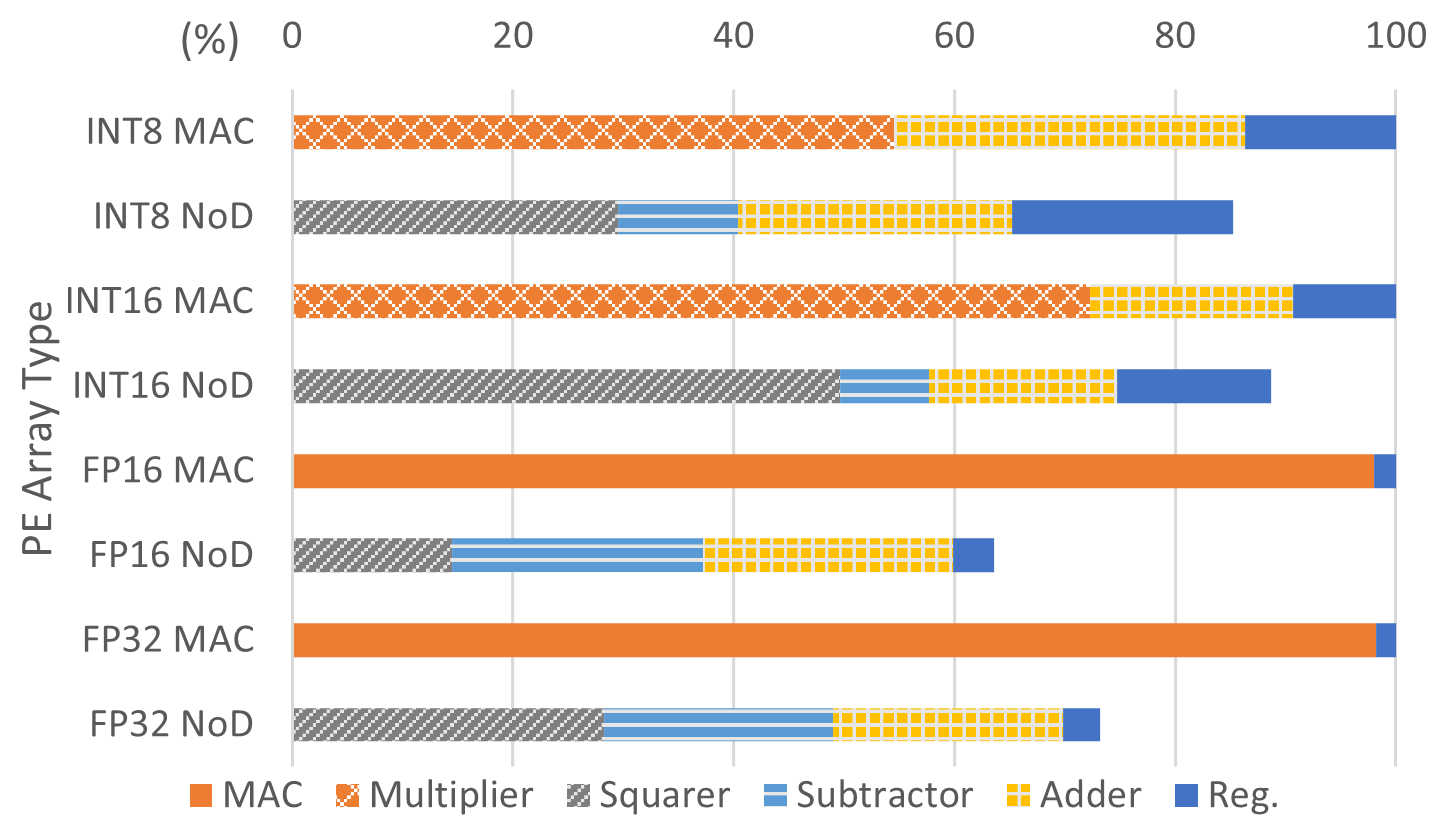}
\caption{Area breakdown of the PE array (normalized to MAC PE array = 100\%).}
\label{Area_breakdown}
\end{figure}

\begin{figure}[tbp]
\centering
\includegraphics[width=1.0\linewidth, keepaspectratio=true]{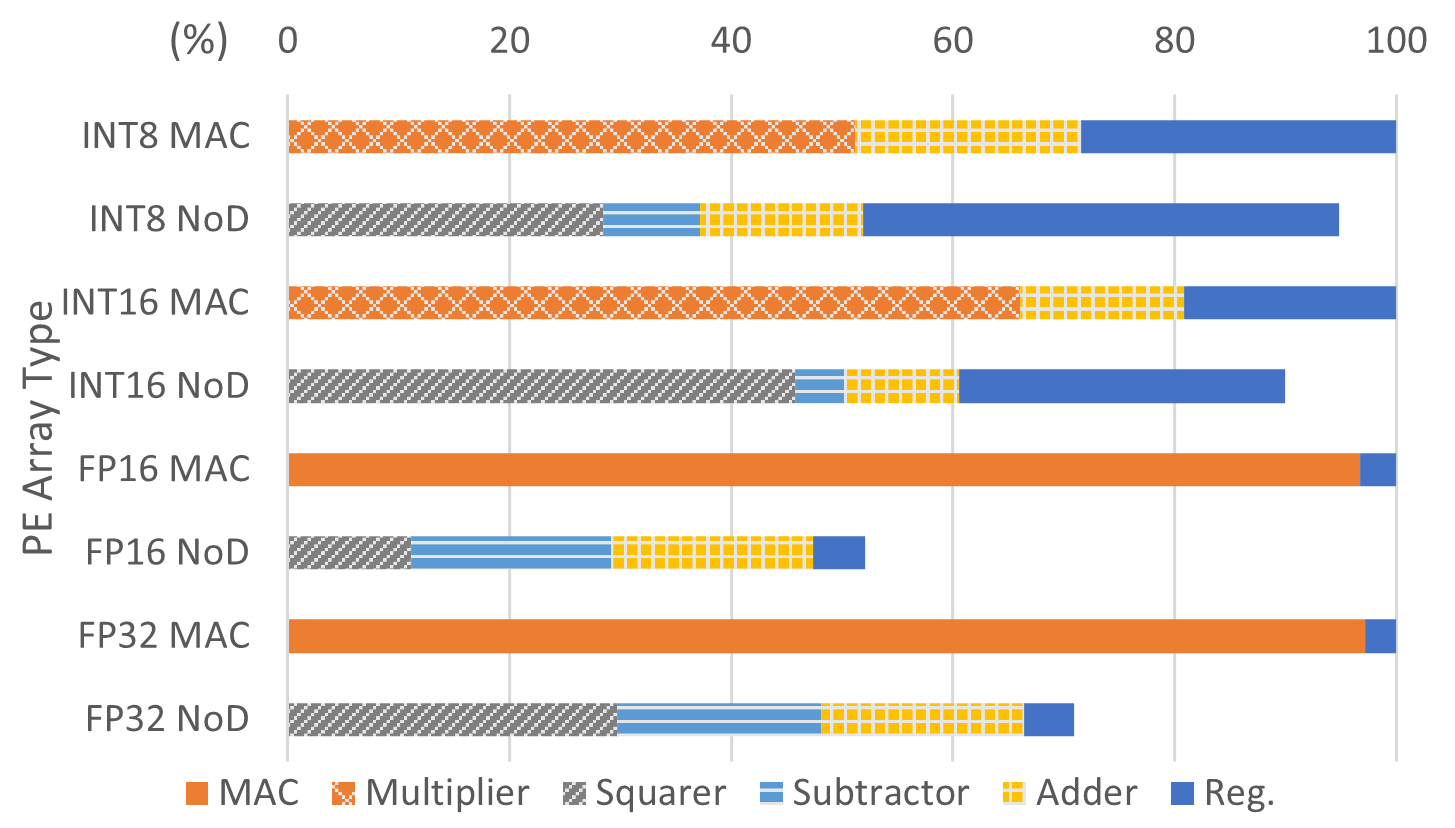}
\caption{Energy breakdown of the PE array (normalized to MAC PE array = 100\%).}
\label{Energy_breakdown}
\end{figure}

\begin{table*}[!htbp]

\caption{\revised{Comparison on IS PE Array and PENDA-CNN Architecture}}

\centering
\begin{tabular}{ccccccccccc}
    \toprule 
    
    Architecture & Precision & PE type & 
    \begin{tabular}[c]{@{}c@{}}Area \\ ($\mu\mathrm{m}^{2}$)\end{tabular} & 
    \begin{tabular}[c]{@{}c@{}}Area\\ Reduction\\ (vs. MAC)\end{tabular} & 
    \begin{tabular}[c]{@{}c@{}}Power\\ (mW)\end{tabular} & 
    \begin{tabular}[c]{@{}c@{}}Normalized\\ Energy\\ ($\mathrm{mW\cdot ns}$)\end{tabular} & 
    \begin{tabular}[c]{@{}c@{}}Energy\\ Reduction\\ (vs. MAC)\end{tabular} & 
    \begin{tabular}[c]{@{}c@{}}Clock \\ Period \\ (ns)\end{tabular} & 
    \begin{tabular}[c]{@{}c@{}}Speedup\\ (vs. MAC)\end{tabular} & 
    \begin{tabular}[c]{@{}c@{}}AET \\ Reduction \\ (vs. MAC)\end{tabular} \\ 
    
    \midrule 

    \multirow{4}{*}{IS Array} 
        & \multirow{2}{*}{INT8}  & MAC & 47803 & - & 196 & 92 & - & 0.47 & - & - \\
        &                        & NoD & 44988 & 6\% & 211 & 91 & 1\% & 0.43 & 9\% & 15\% \\ 
        
        \cmidrule(l){2-11} 
        
        & \multirow{2}{*}{INT16} & MAC & 118050 & - & 421 & 252 & - & 0.6 & - & - \\
        &                        & NoD & 106585 & 10\% & 438 & 228 & 10\% & 0.52 & 15\% & 29\% \\ 
    
    \midrule 

    \multirow{4}{*}{CNN} 
        & \multirow{2}{*}{INT8}  & MAC & 2746 & - & 14.1 & 7.9 & - & 0.56 & - & - \\
        &                        & NoD & 1913 & 30\% & 8.5 & 4.1 & 49\% & 0.48 & 17\% & 69\% \\ 
        
        \cmidrule(l){2-11} 
        
        & \multirow{2}{*}{INT16} & MAC & 7303 & - & 28.0 & 19.6 & - & 0.7 & - & - \\
        &                        & NoD & 5751 & 21\% & 21.7 & 12.6 & 36\% & 0.58 & 21\% & 58\% \\ 
    
    \bottomrule 
\end{tabular}
\\[1.0ex]
\footnotesize
\parbox{\textwidth}{
\textit{Note 1:} Results include the hardware overhead of additional accumulators and buffers for partial sum management required by the IS dataflow.

\textit{Note 2:} Integer accumulator bit widths are provisioned according to Section~\ref{sec:PE}: \revised{24-bit MAC versus 26-bit NoD for INT8 IS; 40-bit MAC versus 42-bit NoD for INT16 IS; 20-bit MAC versus 23-bit NoD for INT8 CNN; and 36-bit MAC versus 39-bit NoD for INT16 CNN.}

\textit{Note 3:} Normalized Energy = Power $\times$ Clock Period
}

\label{tab:is_cnn_comparison}
\end{table*}

To further elucidate the source of these area and energy savings, Fig.~\ref{Area_breakdown} and Fig.~\ref{Energy_breakdown} present normalized area and energy breakdowns of the PE array components. With the baseline MAC-based PE array set to 100\%, the results reveal that the combined area and energy consumption of the subtractor and squarer in the NoD-based PE are lower than those of the single multiplier in the standard MAC-based PE. This architectural simplification is the primary driver for the observed area and energy reductions across all precisions.


\begin{table*}[]

\caption{Comparison with State-of-the-Art PE Array Architectures}

\centering
\begin{tabular}{clcccccccc}
    \toprule 
    
    Precision & PE Array Type & 
    \begin{tabular}[c]{@{}c@{}}Area \\ ($\mu\mathrm{m}^{2}$)\end{tabular} & 
    \begin{tabular}[c]{@{}c@{}}Area\\ vs.\\ MAC\end{tabular} & 
    \begin{tabular}[c]{@{}c@{}}Power\\ (mW)\end{tabular} & 
    \begin{tabular}[c]{@{}c@{}}Normalized\\ Energy\\ ($\mathrm{mW\cdot ns}$)\end{tabular} & 
    \begin{tabular}[c]{@{}c@{}}Energy\\ vs. \\ MAC\end{tabular} & 
    \begin{tabular}[c]{@{}c@{}}Clock \\ Period \\ (ns)\end{tabular} & 
    \begin{tabular}[c]{@{}c@{}}Clock \\ Period \\ vs. MAC\end{tabular} & 
    \begin{tabular}[c]{@{}c@{}}AET \\ vs. \\ MAC\end{tabular} \\ 
    
    \midrule 

    \multirow{4}{*}{INT8} 
        & MAC OS PE Array (Baseline)   & 44607 & - & 180 & 90 & - & 0.5 & - & - \\
        & Winograd FFIP (Original)     & 41930 & $-6\%$ & 175 & 84 & $-6\%$ & 0.48 & $-4\%$ & $-16\%$ \\
        & Winograd FFIP (Full Bits)    & 49066 & $+10\%$ & 193 & 96 & $+7\%$ & 0.5 & $0\%$ & $+18\%$ \\
        & NoD-based OS PE Array (Ours) & 38010 & $-15\%$ & 194 & 85 & $-5\%$ & 0.44 & $-14\%$ & $-29\%$ \\ 
    
    \midrule 

    \multirow{4}{*}{INT16} 
        & MAC OS PE Array (Baseline)   & 108639 & - & 376 & 241 & - & 0.64 & - & - \\
        & Winograd FFIP (Original)     & 94907 & $-13\%$ & 344 & 210 & $-13\%$ & 0.61 & $-5\%$ & $-28\%$ \\
        & Winograd FFIP (Full Bits)    & 117129 & $+8\%$ & 394 & 256 & $+6\%$ & 0.65 & $+2\%$ & $+16\%$ \\
        & NoD-based OS PE Array (Ours) & 96385 & $-11\%$ & 402 & 217 & $-10\%$ & 0.54 & $-19\%$ & $-33\%$ \\ 
    
    \bottomrule 
\end{tabular}
\\[1.0ex]
\footnotesize
\parbox{\textwidth}{
\textit{Note 1:} Positive percentages ($+$) indicate an increase relative to the baseline, while negative percentages ($-$) indicate a reduction (improvement).

\textit{Note 2:} Results for FFIP include the hardware overhead of additional accumulators and buffers required for partial sum management.
}

\label{tab:sota_comparison}
\end{table*}

Table \ref{tab:is_cnn_comparison} extends this analysis to the IS dataflow and the specialized CNN dataflow (detailed in Section \ref{sec:DISCOS CNN}). These comparisons validate that the advantages of PENDA are robust across different architectural configurations and workloads.
For the IS array, PENDA demonstrates consistent AET reductions of 15\% (INT8) and 29\% (INT16). \revised{Similarly, for the CNN architecture, PENDA achieves comparable gains, reducing AET by 69\% (INT8) and 58\% (INT16).} This consistency across OS, IS, and CNN configurations reinforces the versatility of the proposed method, confirming its applicability to a wide range of deep learning accelerators.



\subsection{Comparison with State-of-the-Art PE Array Architecture}

To further evaluate the effectiveness of PENDA, we compare it against the state-of-the-art Winograd FFIP~\cite{pogue2023fast} PE array architecture. To ensure a fair comparison, we reproduced the FFIP array using the same process environment and design flow as our own method.

The FFIP algorithm is defined in~\cite{pogue2023fast} by the following core equations:
\begin{equation}
\label{eq:ffip_main}
c_{i,j} = \sum_{k=1}^{K/2} g_{i,2k-1}^{(j)} \cdot g_{i,2k}^{(j)} - \alpha_{i} - \beta_{j}
\end{equation}
where $g$ is a recursive accumulation defined as:
\begin{equation}
\label{eq:ffip_g}
g_{i,k}^{(j)} = 
\begin{cases} 
      a_{i,2k} + y_{2k-1,j} & \text{for } j=1, k \text{ is odd} \\
      a_{i,2k-1} + y_{2k,j} & \text{for } j=1, k \text{ is even} \\
      g_{i,k}^{(j-1)} + y_{k,j} & \text{for } j > 1 
\end{cases}
\end{equation}

During reproduction, our analysis of \eqref{eq:ffip_g} and the corresponding datapath identified that the $g$ value is continuously accumulated with $y$ values along the systolic array's $j$-dimension. However, the bit-width assigned to the $g$ register in the original work~\cite{pogue2023fast} (defined as $w+1$) is insufficient to guarantee that this accumulation chain will not overflow under worst-case conditions. The original paper~\cite{pogue2023fast} does not provide an analysis for this potential overflow risk.

Therefore, to conduct a fair and complete comparison, we evaluated two versions of FFIP, as shown in Table \ref{tab:sota_comparison}:
\begin{enumerate}
    \item FFIP (Original)~\cite{pogue2023fast}: This version faithfully reproduces the bit-widths ($w+1$) disclosed in the original work~\cite{pogue2023fast}. While it appears superior in resource utilization, it fails to guarantee computational correctness and is susceptible to overflow.
    \item FFIP (Full Bits): This is our corrected version, which expands the bit-widths of the $g$ registers and associated adders to provide the necessary headroom to prevent overflow under all conditions.
\end{enumerate}

The results in Table \ref{tab:sota_comparison} clearly expose a key implementation bottleneck of the FFIP architecture: to guarantee computational correctness, the required hardware resources increase substantially, causing its overall AET efficiency to become even worse than the traditional MAC baseline (e.g., AET increases by 18\% and 16\% for INT8 and INT16, respectively).

In contrast, PENDA, which incorporates sufficient headroom (+2 bits) for the accumulator of NoD-based PE by design (as detailed in Section \ref{sec:PE}), inherently ensures overflow-free operation. Consequently, PENDA achieves superior hardware efficiency and performance (e.g., 29\% and 33\% AET reduction for INT8 and INT16, respectively) significantly better than both the MAC baseline and the FFIP architecture, while guaranteeing computational integrity.

\subsection{System-Level Comparison}
\label{System-Level Comparison}



\revised{
Finally, to demonstrate that the component-level advantages of PENDA translate effectively to full-system designs, we compare the MAC- and PENDA-based implementations of two INT8 accelerator configurations for ViT-B and ResNet-18, as defined in Section~\ref{sec:System Architecture}. The comparisons in Tables~\ref{tab:system_comparison} and~\ref{tab:system_comparison_CNN} account for all system-level components, including PE arrays, SRAM, and PENDA-specific overheads.

The results show that the component-level gains translate directly into system-level improvements. As shown in Table~\ref{tab:system_comparison}, the PENDA-based ViT-B accelerator achieves a 13.6\% speedup and an 18\% overall AET reduction. For ResNet-18, Table~\ref{tab:system_comparison_CNN} shows that the PENDA-based CNN accelerator achieves a 16.7\% speedup and a 59\% AET reduction while reducing both area and total on-chip energy. These results demonstrate that the system-level benefits of PENDA extend beyond Transformer workloads to CNN accelerators.

}

\begin{table}[tbp]

\caption{System-Level Comparison for ViT-B Accelerator}

\centering
\begin{tabular}{lccccc}
    \toprule 
    
    Architecture & 
    \begin{tabular}[c]{@{}c@{}}Area\\ ($\mu\mathrm{m}^{2}$)\end{tabular} & 
    \begin{tabular}[c]{@{}c@{}}Power\\ (mW)\end{tabular} & 
    \begin{tabular}[c]{@{}c@{}}Energy\\ ($\mu\mathrm{J}$)\end{tabular} & 
    Speedup & 
    \begin{tabular}[c]{@{}c@{}}AET\\ Red.\end{tabular} \\ 
    
    \midrule 

    MAC (Baseline) & 408907 & 806 & 611 & - & - \\
    PENDA (Ours)   & 390676 & 878 & 596 & 13.6\% & 18\% \\
    
    \bottomrule 
\end{tabular}
\\[1.0ex]
\footnotesize
\parbox{\linewidth}{
\textit{Note:} Reported energy consumption represents the total on-chip energy required to compute a single ViT-B encoder layer.
}

\label{tab:system_comparison}
\end{table}

\begin{table}[htbp]

\caption{\revised{System-Level Comparison for Resnet-18 Accelerator}}

\centering
\begin{tabular}{lccccc}
    \toprule 
    
    Architecture & 
    \begin{tabular}[c]{@{}c@{}}Area\\ ($\mu\mathrm{m}^{2}$)\end{tabular} & 
    \begin{tabular}[c]{@{}c@{}}Power\\ (mW)\end{tabular} & 
    \begin{tabular}[c]{@{}c@{}}Energy\\ ($\mu\mathrm{J}$)\end{tabular} & 
    Speedup & 
    \begin{tabular}[c]{@{}c@{}}AET\\ Red.\end{tabular} \\ 
    
    \midrule 

    MAC (Baseline) & 789809 & 3189 & 1133 & - & - \\
    PENDA (Ours)   & 614409 & 2261 & 694 & 16.7\% & 59\% \\
    
    \bottomrule 
\end{tabular}
\\[1.0ex]
\footnotesize
\parbox{\linewidth}{
\textit{Note:} Reported energy consumption represents the total on-chip energy required to compute a single image.
}

\label{tab:system_comparison_CNN}
\end{table}

\newpage
\section{Conclusion} 
\label{chapter:conclusion}

\revised{

This paper presented PENDA, which replaces multiplier-dominated MAC units with efficient NoD units while maintaining exact computation. Evaluations demonstrate that NoD-based PE arrays reduce AET by 29\%, 33\%, 72\%, and 53\% for INT8, INT16, FP16, and FP32, respectively, compared with MAC-based arrays, and outperform the evaluated state-of-the-art exact PE-array architecture. These benefits are obtained through a PE-level arithmetic redesign rather than a system-level architectural overhaul. Therefore, PENDA can be applied to GEMM-dominated layers across different models, PE-array organizations, and dataflows with only small system-level overhead. Future work will further explore its integration with emerging low-bit formats. 

}

\bibliographystyle{IEEEtran}

\bibliography{bib/ieeeBSTcontrol,bib/thesis}

\begin{IEEEbiography}[{\includegraphics[width=1in,height=1.25in,clip,keepaspectratio]{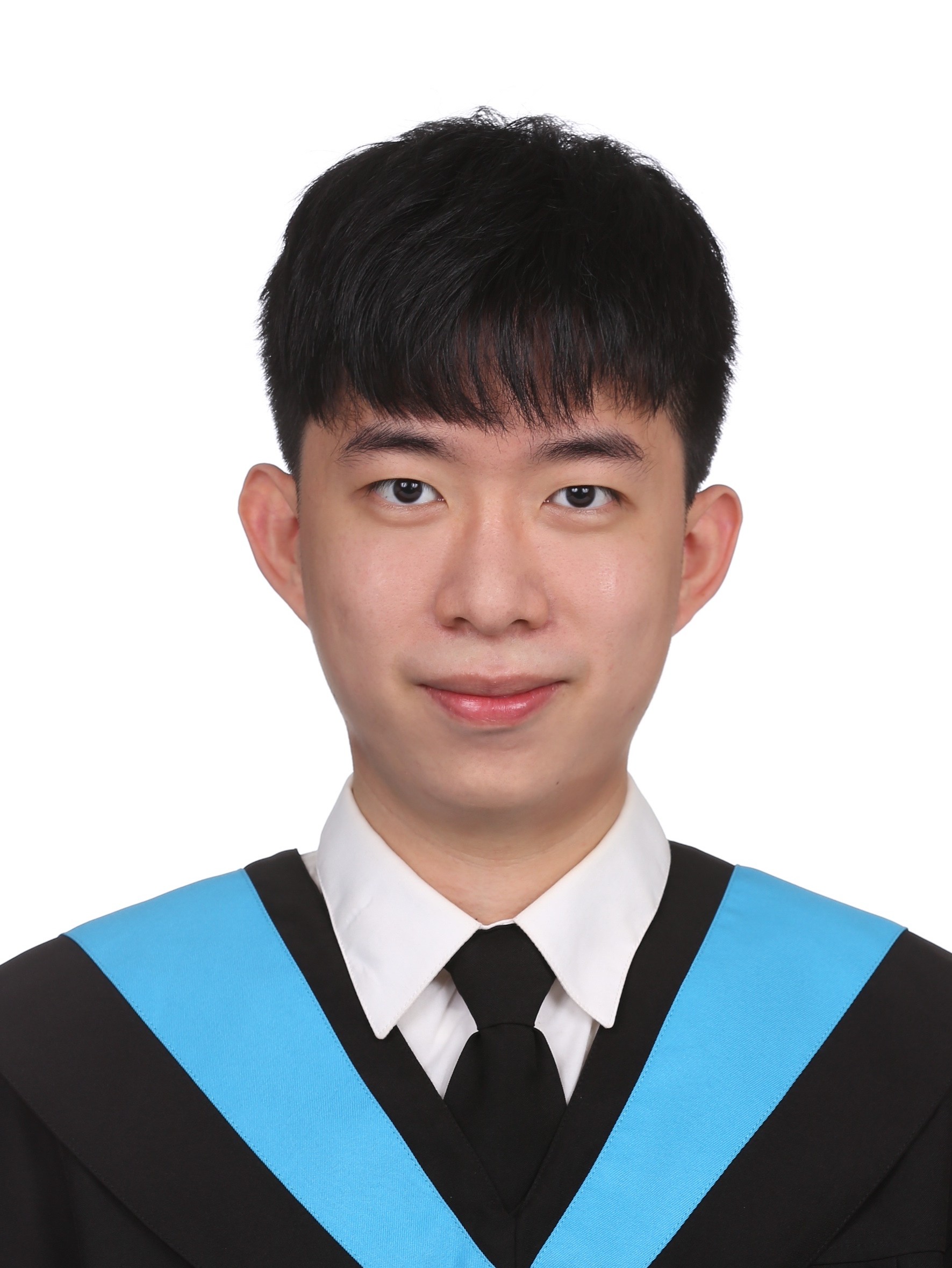}}]{Kai-Chieh Hsu}
received the B.S. degree in electronics engineering from National Yang Ming Chiao Tung University, Hsinchu, Taiwan, in 2022. He is currently pursuing the Ph.D. degree in electronics engineering with National Yang Ming Chiao Tung University. His research interests include AI accelerators for Transformers and CNNs, and VLSI design.

\end{IEEEbiography}

\begin{IEEEbiography}[{\includegraphics[width=1in,height=1.25in,clip,keepaspectratio]{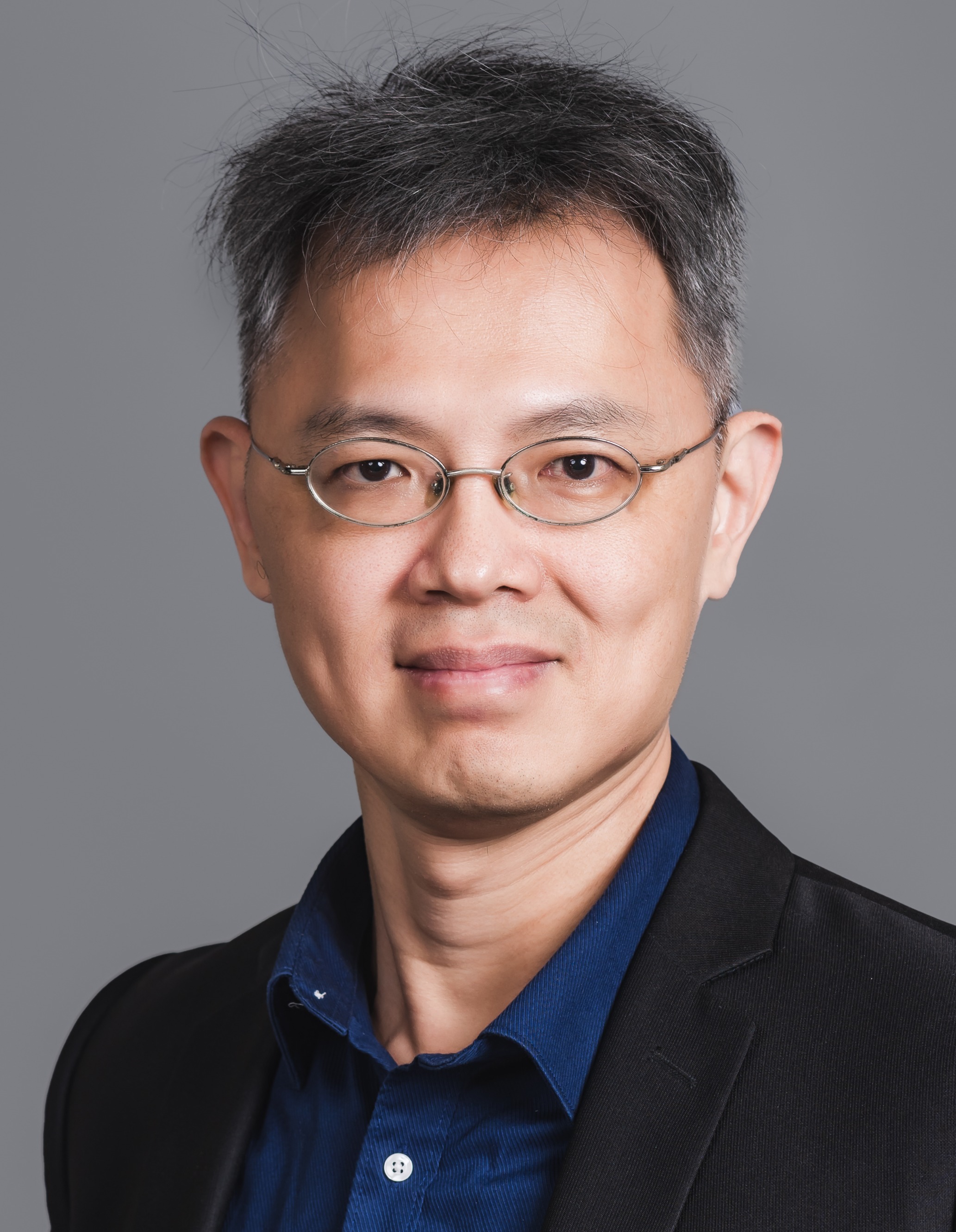}}]{Tian-Sheuan Chang}
	(S’93–M’06–SM’07)
	received the B.S., M.S., and Ph.D. degrees in electronic engineering from National Chiao-Tung University (NCTU), Hsinchu, Taiwan, in 1993, 1995, and 1999, respectively.

	From 2000 to 2004, he was a Deputy Manager with Global Unichip Corporation, Hsinchu, Taiwan. In 2004, he joined the Department of Electronics Engineering, NCTU (as National Yang Ming Chiao Tung University (NYCU) in 2021), where he is currently a Professor. In 2009, he was a visiting scholar in IMEC, Belgium. His current research interests include system-on-a-chip design, VLSI signal processing, and computer architecture.

	Dr. Chang has received the Excellent Young Electrical Engineer from Chinese Institute of Electrical Engineering in 2007, and the Outstanding Young Scholar from Taiwan IC Design Society in 2010. He has been actively involved in many international conferences as an organizing committee or technical program committee member.
\end{IEEEbiography}

\end{document}